\documentclass[preprint,preprintnumbers,amsmath,amssymb,superscriptaddress,floatfix]{revtex4-2}

\usepackage[dvipdfmx]{graphicx,color} 

\usepackage[T1]{fontenc}
\usepackage{lmodern}
\usepackage{braket}

\usepackage{dcolumn}
\usepackage{bm}
\usepackage{color}
\usepackage{hyperref}
\usepackage{array}

\begin{document}

\title{Loading ultracold atoms onto nonlinear Bloch states and soliton states in bichromatic lattices}
\noindent
\author{Tomotake Yamakoshi}
\affiliation{Institute for Laser Science, University of Electro-Communications, 1-5-1 Chofugaoka, Chofu-shi, Tokyo 182-8585, Japan}
\author{Shinichi Watanabe}
\affiliation{Center for International Programs and Exchange, University of Electro-Communications, 1-5-1 Chofugaoka, Chofu-shi, Tokyo 182-8585, Japan}

\date{\today}

\begin{abstract}   
We simulate and analyze an experimental method of loading interacting ultracold atoms onto nontrivial quantum states such as nonlinear Bloch wave and soliton solutions in a 1-dimensional bichromatic lattice.
Of standard bands, inverted bands, and bands with Dirac-like points permitted by a bichromatic lattice, we consider the case of an inverted band and examine the loading process in terms of nonlinear Bloch waves formed by an aggregate of ultracold atoms described by the mean-field model.
Specifically, we solved the Gross-Pitaevskii equation numerically and found an appropriate standing wave-pulse sequence for the inverted band, which sequence proved to be a suitable protocol for producing soliton solutions.
In addition, we examined the effect of an external potential and dynamical instabilities for the post-loading process.
We also provide an appropriate data set for future experimental realization of our findings.
\end{abstract}

\maketitle

\section{introduction}
\label{sect:Introduction}

Recent advancements in experimental techniques with ultracold atoms, nonlinear optics, and exciton-polariton dynamics provide an excellent platform for studying wide varieties of nonlinear wave phenomena\cite{intro,intro2,intro3}.
The Gross-Pitaevskii equation(GPE) well describes these systems\cite{NLBloch2} even though based on the lowest-order mean-filed approximation.
Therefore, solving this equation expectedly allows us to draw analogies between different systems and investigate nonlinear phenomena in a wide range of parameter spaces.
One of the most intriguing topics is the production and application of soliton solutions, namely, non-dispersive localized wave packets\cite{NLBloch1,NLBloch2}.
An ultracold atomic system is suitable for studying nonlinear dynamics due to its high controllability and accessibility.
One aspect is the dynamical control of external potential by varying the parameters of trapping lasers, magnetic coils, etc.
Moreover, it is possible to alter the strength of atom-atom interaction via Feshbach resonance\cite{Feshbach}.
In the recent studies of ultracold atoms, dynamics of solitons\cite{soliton-exp1,soliton-exp2} and soliton trains\cite{soliton-exp3} were experimentally observed using an appropriate control of external potential to Bose-Einstein condensation(BEC).
These results may pave the way, for instance, for constructing new-types of soliton-based high-precision interferometers\cite{soliton-int}.

Another intriguing topic is nonlinear phenomena in the presence of a periodic potential\cite{intro}.
The nonlinearity modifies the energy-band structure, namely a general consequence of the periodic potential so that the now-well-known loop structure appears at a band edge\cite{NLBloch1,NLBloch2}.
In addition to this, the nonlinearity supports spatially localized solutions whose chemical potential lies between two allowed bands.
They are referred to as ``bandgap solitons''\cite{soliton-the1,soliton-the2}.
Some of these solutions are known to be described in terms of nonlinear Bloch waves(NBW)\cite{NLB-Soliton1}.
Using periodic structures such as polariton condensation in a one-dimensional array of buried mesa traps\cite{NBW-pol}, propagating light in a periodic Kerr nonlinear medium\cite{intro2}, and so forth, the NBWs are experimentally realized.
In the nonlinear optics, time-periodic potentials are applied, and the solitons were measured in time domain\cite{optical-soliton}.
In the case of an ultracold atomic system, a spatially periodic potential such as an optical lattice(OL)\cite{intro} is applied by the interference of counter-propagating lasers.
One of the significant advantages of such a system is that the periodic potential can be tuned in time with ease.
This technique allows variation of the band structure, enabling the loading of atoms to a specific state.
Indeed, in a recent experiment with a 2-dimensional checkerboard lattice controlled by an acceleration of the atoms and staggered shift of the lattice, atoms were loaded to an NBW state located at the top of the loop in the ground band to observe its decay rate\cite{loop-exp}.

Ability to produce a specific target state coherently and efficiently is a prerequisite for achieving further applications.
From a theoretical point of view, several techniques have been suggested, e.g., a fast-forward process\cite{FASTF}, a mode-matching technique\cite{MODEM} and so forth.
However, these techniques generally require some complex coherent control so that it is exceedingly challenging for experimental implementation.
One of the very promising techniques applicable to the ultracold atomic lattice system is the ``standing-wave pulse sequence method''\cite{Beijing-1,Beijing-2,Beijing-5}, also known as the Bang-Bang (on-off) control\cite{KDs}.
Basically, it turns the optical lattice on and off with appropriate time intervals, which is a sequence of square pulses until the residual BEC wave function coincides with the targeted Bloch wave.
An experimental group in Beijing applied this technique to 1-, 2-, and 3-dimensional systems and demonstrated that the loading process is very efficient.
Interestingly, they demonstrated recently that the same technique goes beyond just preparing the initial state, but it would be capable of state-to-state manipulation in the manner of the Ramsey interferometry\cite{Beijing-6}.

In our previous paper\cite{my-4}, we theoretically extended the experimental work of the standing wave pulse method\cite{Beijing-1,Beijing-2,Beijing-5} to the bichromatic lattice system\cite{exp-so} in the linear regime.
We also analyzed the selection rule and numerically optimized the time sequence using experimentally realizable parameters.
Furthermore, we found that relatively weak nonlinear interaction (much smaller than lattice height) does not alter the selection rule and post-loading dynamics since the band dispersion is not dramatically altered under the given nonlinearity condition.
Effects of nonlinear interaction thus remained to be investigated under stronger conditions.

In this paper, we treat a straightforward procedure for loading atoms onto excited NBWs and solitons via the standing-pulse wave sequence in the bichromatic lattice.
We consider the region where the interaction term does not go so far as to form the loop in the band structure.
First, we analyze selection rules for transitions in the nonlinear system based on numerically obtained NBW bands. 
Since the NBWs are not orthogonal to each other, we numerically solve the time-dependent Gross-Pitaevskii equation(TD-GPE) to examine the selection rules.
We will find that it is possible to load atoms onto soliton solutions in the periodic potential once appropriate parameters are obtained.
Subsequently, we examine post-loading effects of the external trapping harmonic potential and reveal that the dynamics reflect the nonlinear band dispersion.
Finally, we discuss the dynamical instability using the method of linear stability analysis.

In passing, we note that the main observables in this paper are density distributions in position and momentum space, which are measurable in experiments.
Besides fundamental techniques, our procedure requires only the control of the interaction strength that appears in GPE.
Furthermore, experimentally necessary parameters are taken from the papers of the Beijing group\cite{Beijing-2,Beijing-4}; thus, this paper contributes to future experiments without any further complex coherent control.

The paper is organized as follows.
Sect.\ref{sect:system} outlines the theoretical model-system and shows nonlinear band dispersion.
Sect.\ref{sect:loading} discusses the selection rule of the loading process in terms of nonlinear Bloch waves.
Sect.\ref{sect:numerics} analyzes the numerical result of the loading process and post-loading dynamics.
Sect.\ref{sect:conclusions} concludes the paper.
Technical details are supplemented in Appendices.

\section{Mathematical Definition and basic features of the System}
\label{sect:system}

In the experimental papers\cite{Beijing-1,Beijing-2,Beijing-3,Beijing-4}, the loading process begins with BEC in a 3D harmonic trap.
The dynamics is 3D and non-separable since the inter-atomic interaction is in effect.
However, here we consider the case where the harmonic confinement is much tighter in the $y$- and $z$-directions than in $x$-direction so that the system can be treated as quasi-1D.
Optimally parametrized 1D OL pulses are applied, and the momentum distributions of the post-loading process are measured after the band mapping process.

Here, we solve the 1D TD-GPE\cite{NLBloch2} for interacting bosonic atoms in the bichromatic OL.
Some notations and techniques used in this paper are available in our numerical studies presented in Ref.'s\cite{my-1,my-3,my-4}.
The TD-GPE reads
$$ i \hbar \frac{d}{dt} \Psi(x') = \left[ -\frac{\hbar^2}{2m_{a}}\frac{\partial^2}{\partial x'^2}+ \alpha(t) 
\{ V_1 \sin^2 (k_{r}x') +  V_2 \sin^2 (2 k_{r} x') \} + \frac{1}{2}m_a\omega_0^2 x'^2 + g_{1D}N |\Psi(x')|^2 \right] \Psi(x')$$
where $V_1$ is the height of the optical lattice with the period of $\lambda/2$, $V_2$ is
with the period of $\lambda/4$, $\alpha(t)$
equals 1 during an on-duty cycle, otherwise it is 0, $\omega_0$ is
the frequency of the harmonic trap, $N$ is the number of total atoms and $g_{1D}$ parametrizes
the effective atom-atom interaction obtained by contracting the 3D trap to 1D.
We use recoil energy $E_r=\hbar^2 k_{r}^2/2 m_{a}$ as the unit of energy, recoil momentum $k_{r}=2\pi/\lambda$ as the unit of (quasi-)momentum, lattice constant $\lambda/2$ as the unit of length and rescaled time $t=E_r t'/\hbar$ as the unit of time.
Here $\hbar$, $\lambda$ and $m_{a}$ correspond to the Planck constant, the wavelength of the laser used for generating OL, and the atomic mass, respectively.
Rescaling the TD-GPE, we get
\begin{eqnarray}
i \frac{d}{dt} \Psi(x) = \left[ -\frac{\partial^2}{\partial x^2}+ \alpha(t) \{ s_1 \sin^2 (x) +s_2 \sin^2 (2x) \} + \nu x^2 + g |\Psi(x)|^2 \right] \Psi(x)
\label{eq:re-ham}
\end{eqnarray}
where $x$, $s_1$, $s_2$ and $g$ denote $x=k_{r}x'$, $s_1=V_1/E_r$, $s_2=V_2/E_r$ and $g=g_{1D}N/E_r$ respectively.
The atom treated here is $^{87}Rb$ and typically $g$ ranges from $10^{-5}$ to $1$\cite{G1D,my-4}. 
The other parameters are the same as in Ref.\cite{Beijing-1}.
We note that $g$ could be made to take on any value by exploiting the Feshbach resonance\cite{Feshbach}.
We limit ourselves to the regime where $g \leq 8$ is comparable to the kinetic energy of the 2nd excited band in this paper.

\begin{figure}[htbp]
 \begin{center}
 \includegraphics[width=8cm]{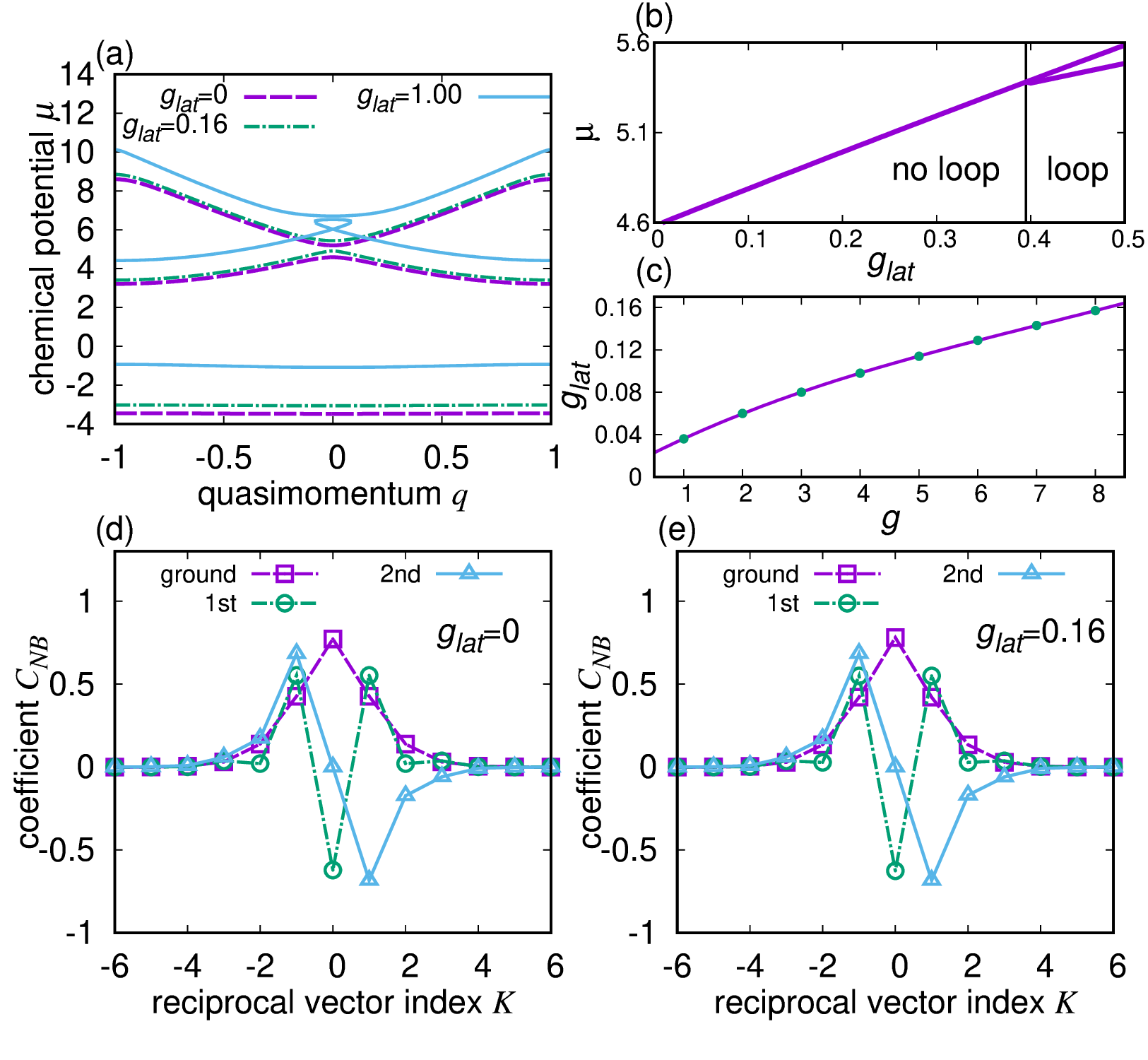}
 \end{center}
 \caption{
(a) The band structure as a function of quasimomentum $q$ for $g_{lat}=0$(linear, purple dashed), $g_{lat}=0.16$(green chain) and $g_{lat}=1$(light blue solid)with $s_1=10,s_2=8$.
Chemical potential showed up to the 2nd excited band.
(b) The chemical potential $\mu$ of the nonlinear Bloch solutions as a function of interaction strength $g_{lat}$ at $q=0$ in the range of $4.6<\mu<5.6$.
If the interaction term is sufficiently small, only one eigenvalue appears for a given value of $g_{lat}$  in this range.
The loop appears at a point a little short of $g_{lat}=0.4$.
(c) Numerically estimated interaction strength for the lattice system $g_{lat}$ as a function of $g$.
The green circles are the numerically obtained result, and the purple solid is a fitted curve $g_{lat}=8.8\times10^{-3}+2.9\times10^{-2}g-2.2\times10^{-3}g^2+1.0\times10^{-4}g^3$.
(d) and (e) corresponding to the first three Bloch coefficients as a function of $K$ at $q=0$ for $g_{lat}$=0 and 0.16, respectively.
The odd excited and the even excited bands are pairwise inverted.
Thus the 1st excited has even symmetry, and the 2nd has odd symmetry.
In this region, the nonlinear interaction slightly alters the coefficients but does not modify the symmetry.
 }
 \label{fig:LBS}
\end{figure}

The nonlinear Bloch waves $\{\phi_{NB}(n,q,x)\}$ formed by the bichromatic OL alone satisfies the time-independent equation, namely, 
\begin{eqnarray}
\left\{ -\frac{\partial^2}{\partial x^2} +  s_1 \sin^2 (x) +s_2 \sin^2 (2x) + g_{lat} | \phi_{NB}(n,q,x) |^2 \right\} \phi_{NB}(n,q,x) = \mu \phi_{NB}(n,q,x) 
\label{eq:NBLE}
\end{eqnarray}
subject to the normalization condition $\int_{-\pi/2}^{\pi/2} | \phi_{NB}(n,q,x) |^2 dx = \pi$.
Here $g_{lat}$ and $\mu$ correspond to the nonlinear interaction strength for NBWs and the chemical potential, respectively.
Same as in the linear case\cite{SSP}, the NBW can be expanded as
\begin{equation}
\phi_{NB}(n,q,x) = e^{iqx} \sum_{K} C_{NB}(n,q,K) e^{2iKx}
\end{equation}
where $q$ is the quasimomentum, $n \in \mathbb {N}$ is the band index, $K \in \mathbb {Z}$ is the reciprocal vector index.
From this, the coefficient $C_{NB}(n,q,K)$ can be obtained by solving the nonliner reccurent formula under the condition of $\sum_K |C_{NB}(n,q,K)|^2=1$\cite{NLBloch1,NLBloch2,NLB-Soliton1}.
Fig.\ref{fig:LBS}(a) shows the first three band dispersions with $s_1=10,s_2=8$ with $g_{lat}=0$(linear), $0.16$ and $1$.
The trend of energy band dispersion is the same as a linear one if $g_{lat}=$0.16, except the interaction shifts the dispersion higher.
Normally, the odd-excited band's coefficients have odd-symmetry in $K$-space and vice versa.
However, the parity of wave function of the 1st and 2nd excited bands are inverted.
Fig .\ref{fig:LBS}(d) and (e) show the coefficient $C_{NB}(n,q,K)$ of the first three bands at $q=0$ with $g_{lat}=0$ and $0.16$, respectively.
In the linear limit, this parity inversion occurs with a change in the band structure: as $s_2$ increases, the gap between the first and second excited bands becomes smaller, and at the critical point $s_2=(s_1/4)^2=6.25$, the gap disappears.
For bigger $s_2$, the gap becomes larger and the parity of the wave functions of the first and second excited bands is inverted.
See Ref.\cite{my-4} for more detailed discussion.

Because the NBWs are not mutually orthogonal, non-zero overlap integrals occur between different momentum components of the target and unwanted states.
Therefore, it appears worth examining the loading process numerically with TD-GPE.
Furthermore, the nonlinear term is known to give rise to additional solutions other than usual Bloch solutions.
One is the solutions in momentum space corresponding to loop structures that appear at band edges.
Fig.\ref{fig:LBS}(b) shows interaction strength dependency of the 1st excited band structure at $q=0$.
The interaction strength gradually modifies the structure, and the loop appears at around $g_{lat}=0.4$.
For instance, we plotted a band structure with $g_{lat}=1$, which has a loop structure at $q=0$ in the first excited band(Fig.\ref{fig:LBS}(a)).
Typically, the top of the loop and the bottom of the 2nd excited band become flat(See Appendix.\ref{sect:loop} for the detail).
In addition to the NBWs, spatial soliton solutions having their chemical potential lying in a linear bandgap are also allowed.
These solutions are called the ``gap soliton,'' consisting of NBWs with a hyperbolic-secant-type envelope typically.
Therefore, this paper mainly aims to investigate how the loading process is affected by these additional solutions.
See Refs.\cite{NLBloch1,NLBloch2,NLB-Soliton1,NLBloch-bichro} for further detail about the loop structures and the solitons.

As seen in Eq.(\ref{eq:NBLE}), $g_{lat}$ corresponds to the nonlinear interaction at a specified single-site; thus, the theory works perfectly if the wave function is thoroughly uniform (or perfectly periodic) whole over the spatial region.
However, in our case, the simulation begins with the ground state of the harmonic system; therefore, the wave function is nonuniform and localized around the origin of the harmonic trap.
Here we parametrize the interaction strength of the lattice system $g_{lat}=g\overline{n}/\pi$ by averaged peak site density $\overline{n}$ after the loading process, where
$\overline{n}=\int_{-\pi/2}^{\pi/2} |\Psi(x,\tau_{total})|^2 dx$ and $\Psi(x,\tau_{total})$ is the wave function immediately after the loading.
This argument is the same as in the experiment paper done by Koller {\it et.al.} \cite{loop-exp}.
Here, we use an equation obtained by a fitting for the relationship between $g$ and $g_{lat}$ under the condition of loading atoms onto the 1st excited band with $s_1=10, s_2=8$(Fig.\ref{fig:LBS}(c)).
This relationship is valid when $1 \leq g \leq 8$.
For instance, in our numerical simulation, $g=8$ corresponds to $g_{lat}=0.16$.

\section{Loading process}
\label{sect:loading}

In the previous paper\cite{my-4}, we numerically optimized the loading process to the 1st and the 2nd excited bands with a two-step on- and off- procedure in the linear limit as follows.
We assumed that the atoms are initially installed in the ground state of the harmonic potential so that the initial state is an even symmetric 0-momentum state.
Under this assumption, the time evolution during the on-duty cycle is given by the superposition of linear Bloch waves at $q=0$.
And, the off-duty cycle changes the phase of the discretized momentum state.
Thus, the net result can be regarded as a multi-path interference effect in momentum space.
With a grid spacing of 0.1$\mu$s grid, we obtained through brute force the most appropriate time sequence $(\tau_1,\tau'_1,\tau_2,\tau'_2)$ for $q=0$ with the total time $\tau_{total}=\tau_1+\tau'_1+\tau_2+\tau'_2$ shorter than 100 $\mu$s. 
Here $\tau_n$ and $\tau'_n$ are $n$-th time duration for on- and off-duty cycles.
According to the experimental paper Ref.\cite{Beijing-1}, the total time $\tau_{total}$ is comparable to the scale of the recoil energy, since the unit of time in our definition $t=1$($\hbar/E_r$) corresponds to $50.4\mu$s. 
To verify the availability of the protocol, subsequently, we applied the pulse-sequence method to a weakly interacting system($g=0-1$) by solving TD-GPE, starting with the ground state of the harmonic potential as before.
Finally, we confirmed the protocols which is obtained in the linear limit were still valid for the weakly interacting system by calculating the density distribution after the band mapping process.
Especially, in the previous paper\cite{my-4}, we focused on three sets of parameters in the symmetric(even to even) process, so that we had normal band structure($s_1=10, s_2=5$), band with the Dirac-like point ($s_1=10, s_2=6.25$), and inverted band structure ($s_1=10, s_2=8$).
This paper focuses on the inverted band structure, as shown in Fig.\ref{fig:LBS}.
Throughout this paper, we use the optimized protocol in the linear limit(Table.\ref{tab:two-col}) although the nonlinear term takes place.

As seen in Eq.\ref{eq:re-ham}, a minimum of the OL term is set to coincide with that of the harmonic oscillator; that is, the pulse is symmetric under $x\rightarrow -x$.
When this pulse acts on the 0-momentum initial state, which is symmetric, the solution evolves in times, maintaining the initial symmetry.
On the other hand, loading atoms onto anti-symmetric solutions require an additional phase term in the OL pulse to break the symmetry of the initial state.
Now consider, we apply the phase shifted OL pulse to the system, namely
$$s_1 \sin^2(x+\phi_{11}) + s_2 \sin^2(2 \{x+\phi_{21}\})$$
for the 1st on-duty cycle, and
$$s_1 \sin^2(x+\phi_{12}) + s_2 \sin^2(2 \{x+\phi_{22}\})$$
for the 2nd one.
Indeed, the Fourier coefficients of the OL other than the constant term $\frac{s_1+s_2}{2}$ is essential in the discussion of symmetry.
If, for example, $\phi_{11}=\phi_{21}=\pi/4$ and  $\phi_{12}=\phi_{22}=\pi/8$, the OL becomes anti-symmetric under $x\rightarrow -x$ as can be seen by shifting the OL potential by a constant $\frac{s_1+s_2}{2}$.
However, here we simply consider $\phi_{11}=\phi_{21}=\phi_{12}=\phi_{22}=\pi/4$ for experimental implementation although the pulse is only partially anti-symmetric.
See Appendix.\ref{sect:optimize} for the detail about the effect of relative phases.
Let us consider the reachable target state with the anti-symmetric loading process at $q=0$ in the linear limit.
When the lattice height $s_2$ is lower than the critical value $(s_1/4)^2$, the linear Bloch wave at $q=0$ of the 1st excited band has odd symmetry in $K$-space, and 2nd has even.
Therefore, the anti-symmetric loading process fails to load atoms onto the 2nd excited band.
Once $s_2$ exceeds $(s_1/4)^2$, the parity of the linear Bloch waves in $K$-space is inverted(Fig.\ref{fig:LBS} (d)), so the atoms cannot be loaded onto the 1st excited band with the anti-symmetric loading process.

In $K$-space, the 1st excited band solution with $g_{lat}=0.16$ behaves almost the same as the linear one (see Fig.\ref{fig:LBS}). 
Therefore, even when the effect of the nonlinear term becomes a little stronger in this case, we try loading the atoms using the same time sequence as in the linear case shown in Table.\ref{tab:two-col}.
Let us note that unlike the linear case, the time evolution cannot be described by an expansion over eigenstates in the nonlinear case.
Therefore, we applied the 4th order split-step Fourier method for time propagation\cite{Tannor}.
When the interaction strength is strong enough, a new NBW solution emerges, which corresponds to the loop structure.
In addition to this interesting phenomenon, there appears another nontrivial set of spatial solutions called ``solitons''.
The main focus of this paper is to present a method to load atoms onto NBWs and also onto soliton solutions.
We then discuss the stability of the solitons after the loading.

\begin{table}[h]
\caption{
Optimized loading protocols to the first and the second excited bands with two step on- and off- procedure by using a brute force method with a period of 0.1$\mu$s grid\cite{my-4}.
The lattice height is $s_1=10,s_2=8$.
In this paper we chose relative phase $\phi_{11}=\phi_{21}=\phi_{12}=\phi_{22}=\pi/4$ for anti-symmteric loading for on-duty cycles.
$F$ is the fidelity of the optimization protocol in the linear limit.
See Appendix.\ref{sect:optimize} for more detail.}
\begin{tabular}{c|cccc|c}
\hline\hline
  target   & $\tau_1$   &   $\tau'_1$       &  $\tau_2$    & $\tau'_2$    &  $F$   \\
\hline
1st excited,even        &   13.0       &   22.9            &   2.5        &   30.4      & 0.995 \\
2nd excited,odd         &   33.8       &   34.7            &   14.8       &   4.5       & 0.939 \\
\hline\hline
\end{tabular}
\label{tab:two-col}
\end{table}

\section{Discussions on Numerical Results}
\label{sect:numerics}

This section discusses the numerical result of the loading process with the inverted band structure ($s_1=10,s_2=8$) for the interaction strength $g=1-8$.
In the experiments \cite{Beijing-2,Beijing-5}, the loading efficiency is estimated by observing the post-loading dynamics of the atoms by suddenly freezing the OL in time.
The Faraday imaging method can directly observe the motion of atoms in position space with sufficiently high resolution\cite{Faraday}.
We note that, in our parameter regime, single-site resolved imaging is not necessary to observe the dynamics of wave packets since the wave packets are spatially spread over several sites.
Observing wave packets with single-site resolved imaging would help understand more detailed dynamics and structures of the solitons.
As for the experimental techniques of single-site resolved imaging, refer Ref.'s.\cite{single-atom,Faraday-SS} and references therein.
The band mapping technique is one of the most promising experimental methods that map the quasimomentum of the loaded atoms onto the real momentum domain and reveals the band indices\cite{bmap-n,Hamburg,Aarhus1}.
Therefore, we analyze the density distribution in momentum space as well as in position space, which is obtained by the band mapping procedure, in order to study the post-loading dynamics.
For the numerical implementation, we chose 1.0ms for the whole band mapping process with a decay constant $\gamma=500\mu{\rm s}$ according to Ref.\cite{Beijing-2}.

First, we discuss the post-loading dynamics with only the OL on (Case 1).
We show that the loading process can load atoms onto soliton solutions if the interaction strength is high enough.
Second, we study the case when the harmonic potential is added, and discuss how the wave-packet behaves under the influence of the harmonic trap (Case 2).
Finally, we study the post-loading dynamics when the initial state is not a clean ground state but affected by an artificial noise.
And we discuss the effect of dynamical instabilities (Case 3).
In all the cases listed above, the numerical simulation begins with the ground state $\psi(x)$ in the harmonic trap with $\nu=1.2\times10^{-4}$ (corresponding to $2\pi\times70$Hz) at a certain interaction strength $g$.
And in Case 3, we use the ground state $\psi(x)$ with additional artificial Gaussian noise as the initial state to simulate TD-GPE.

\subsection{Case 1: Post-loading dynamics with only OL}
\label{sect:excitation}

The mean momentum of the initial wave function is 0, and its width in momentum space is around $0.1k_r$.
The loading process essentially preserves quasimomentum since the time required for the loading is short enough.
Therefore, the loaded atoms can appear throughout many bands, but only around $q=0$.
If there is no interaction and if there is no external potential other than OL, the atoms loaded in a particular band spread out on the time scale deriving from the tunneling term.
The dynamics is described by the superposition of various Bloch states around $q=0$ in a specific band, so that the interference of these closely located states leads to the appearance of fringes.
The interference pattern depends on the dispersion of the loaded band.
In other words, measuring the post-loading dynamics with OL allows us to analyze the band dispersion around $q=0$.
Examination of this issue is interesting, particularly in the presence of the Dirac point\cite{Dirac-t,Dirac-e,my-4}, but we present this case in a separate article.

\begin{figure}[htbp]
 \begin{center}
 \includegraphics[width=8cm]{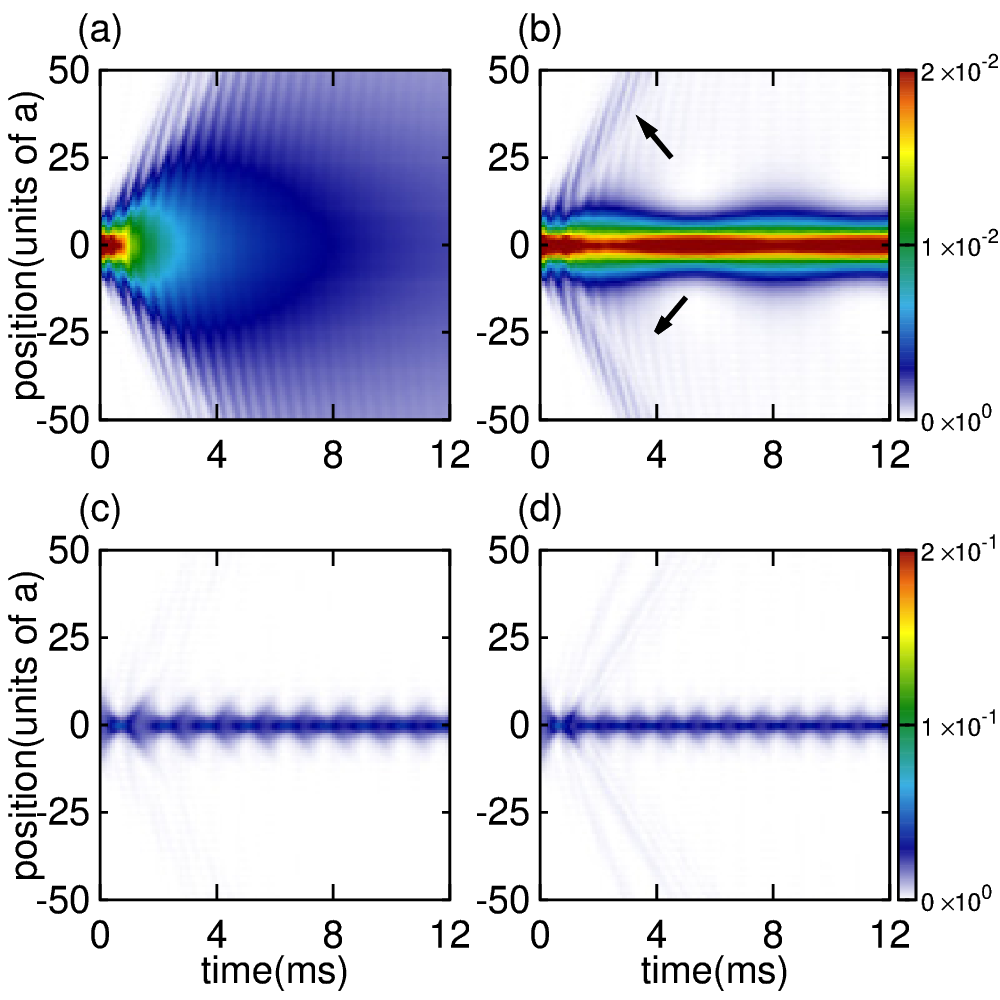}
 \end{center}
 \caption{
Time evolution of the wave packet in position space as a function of holding time with $s_2=8$ for (a)$g=1$, (b)$g=2$, (c)$g=4$ and (d)$g=8$.
In (a), the wave packet spreads out over time due to the tunneling.
When the interaction strength becomes $g\geq 2$, the atoms are loaded to soliton states, and most of the wave packets localize at the origin of the initial state over time.
As the interaction strength $g$ becomes stronger, the frequency of the breathing mode increases and the width of the soliton decreases.
The visible expanding components(faint blue line:black arrow) mainly consist of 1st excited band component in (b).
Similar components appear in (c) and (d), although the density is lower.
 }
 \label{fig:S8-DENSITY}
\end{figure}

\begin{figure}[htbp]
 \begin{center}
 \includegraphics[width=8cm]{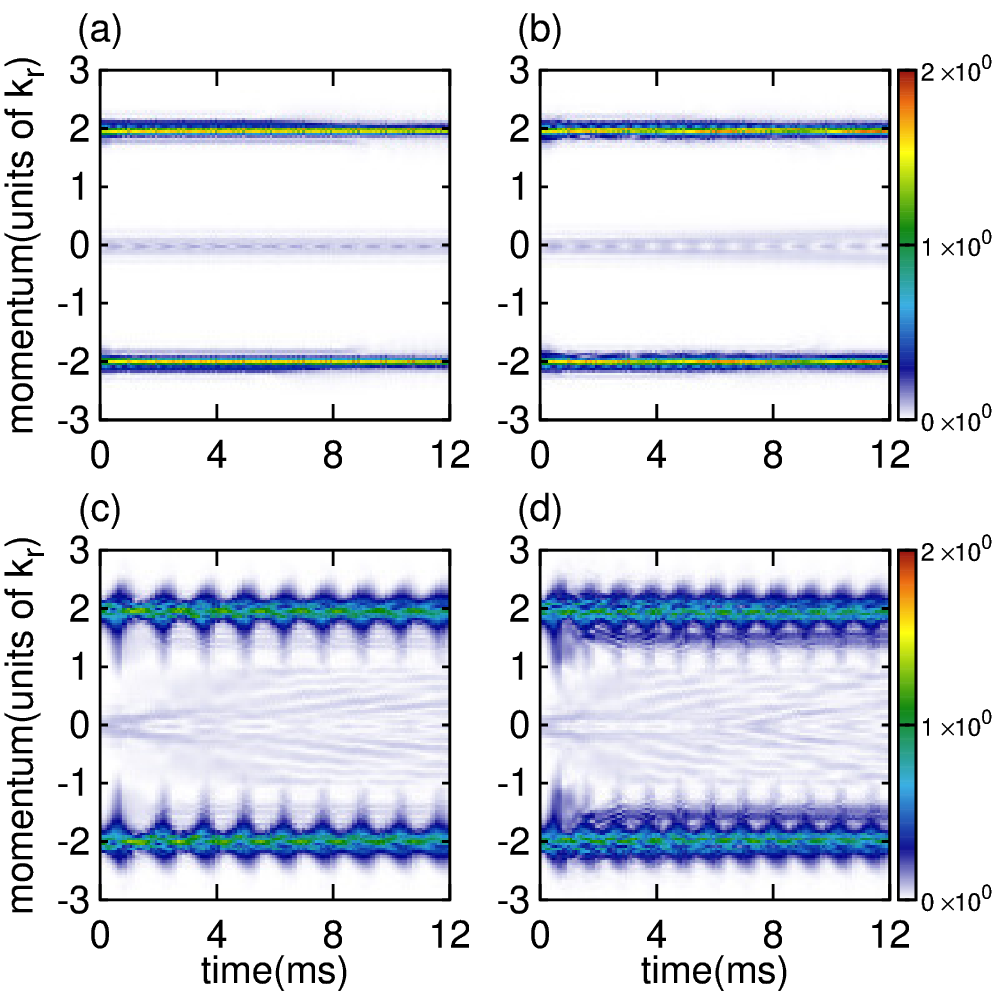}
 \end{center}
 \caption{
 Time evolution of the wave packet in momentum space after the band mapping process as a function of holding time with $s_2=8$ for (a)$g=1$, (b)$g=2$, (c)$g=4$ and (d)$g=8$.
 In all cases, the wave packet is located at around $p=2$.
 In (c) and (d), the oscillation of the breathing mode can be seen clearly.
 }
 \label{fig:S8-MOMENTUM}
\end{figure}

In this subsection, we examine the effect of the interaction term on the post-loading dynamics.
Here, we apply the two-step duty-cycle shown in Table.\ref{tab:two-col}, and then turn on only the OL and keep it on after the loading process.
Fig.\ref{fig:S8-DENSITY} shows the dynamics of the loaded atoms in position space after the loading to the symmetric solution of the 1st excited band(even to even).
And the corresponding momentum distribution after the band mapping process is shown for various cases in Fig.\ref{fig:S8-MOMENTUM}.
When the interaction strength is weak enough (a)$g=1$, the post-loading dynamics directly reflects the local nonlinear band dispersion as shown in Fig.\ref{fig:LBS}.
In position space, the wave packet spreads out over time centered at the origin of the initial state while the momentum distribution(Fig.\ref{fig:S8-MOMENTUM} (a)) is localized at $|p|=2$.
Above $g=2$, the dynamics shows solitonic behavior.
The wave packet is localized around the origin of the initial state and shows a breathing oscillation\cite{soliton-load,soliton-the2} in position space without significant spreading (Fig.\ref{fig:S8-DENSITY} (b)-(d)).
On the other hand, in momentum space, it remains almost unchanged, and the atoms are localized around $|p|=2$ (Fig.\ref{fig:S8-MOMENTUM} (b)-(d)).
In terms of band dispersion, an immobile wave packet corresponds to a flat-like dispersion.
The loop top dispersion becomes flat-like if the interaction strength is high enough $g_{lat} \ge 0.4$.
However, even without such shape deformation of the band dispersion, localized wave packets appear(see Fig.\ref{fig:S8-DENSITY}(b):$g_{lat}=0.06$).
The results suggest that the localization of the wave packet is not merely due to the band dispersion but to the appearance of soliton solutions.
We also evaluate the loading rate around $p=|m|(m \in \mathbb{N})$ by integrating the wavefunction in the momentum space $R_{m}=\left(\int_{m-0.5}^{m+0.5} |\psi(p)|^2 dp + \int_{-m-0.5}^{-m+0.5} |\psi(p)|^2 dp \right)/\int_{-\infty}^{\infty} |\psi(p)|^2 dp$.
Here $\psi(p)$ is obtained by applying the band mapping procedure to the wave function immediately after the loading $\psi(x,\tau_{total})$.
For $m=0$, the range of integration is simply $p=-0.5$ to $0.5$.
Table.\ref{tab:POP} shows the results for $g=1$, 2, 4 and 8.
We note that the loading rate is just the population around the $m$-th band edge.
In order to estimate the band population more precisely, the wave packet must be moved far enough away from the band edges.
Due to the nontrivial nonlinear interaction, the relationship between the loading rate and the interaction strength cannot be expressed by a monotonic function.

\begin{table}[h]
\caption{
The loading rate $R_{m}$ with $g=$1, 2, 4 and 8.
In all cases, $R_{1}$, $R_{3}$ are always smaller than $10^{-3}$.
$R_2$ is about 10\% smaller than the optimized Fidelity $F$ in the linear limit.
}
\begin{tabular}{c|c|c|c|c}
\hline\hline
           &    $g=1$    &    $g=2$      &  $g=4$      &  $g=8$     \\
\hline
 $R_0$     &    0.031    &    0.062      &  0.037     &  0.037      \\
 $R_2$     &    0.855    &    0.929      &  0.876     &  0.888      \\
 $R_4$     &    0.058    &    0.004      &  0.042     &  0.032      \\
\hline\hline
\end{tabular}
\label{tab:POP}
\end{table}

\begin{figure}[htbp]
 \begin{center}
 \includegraphics[width=12cm]{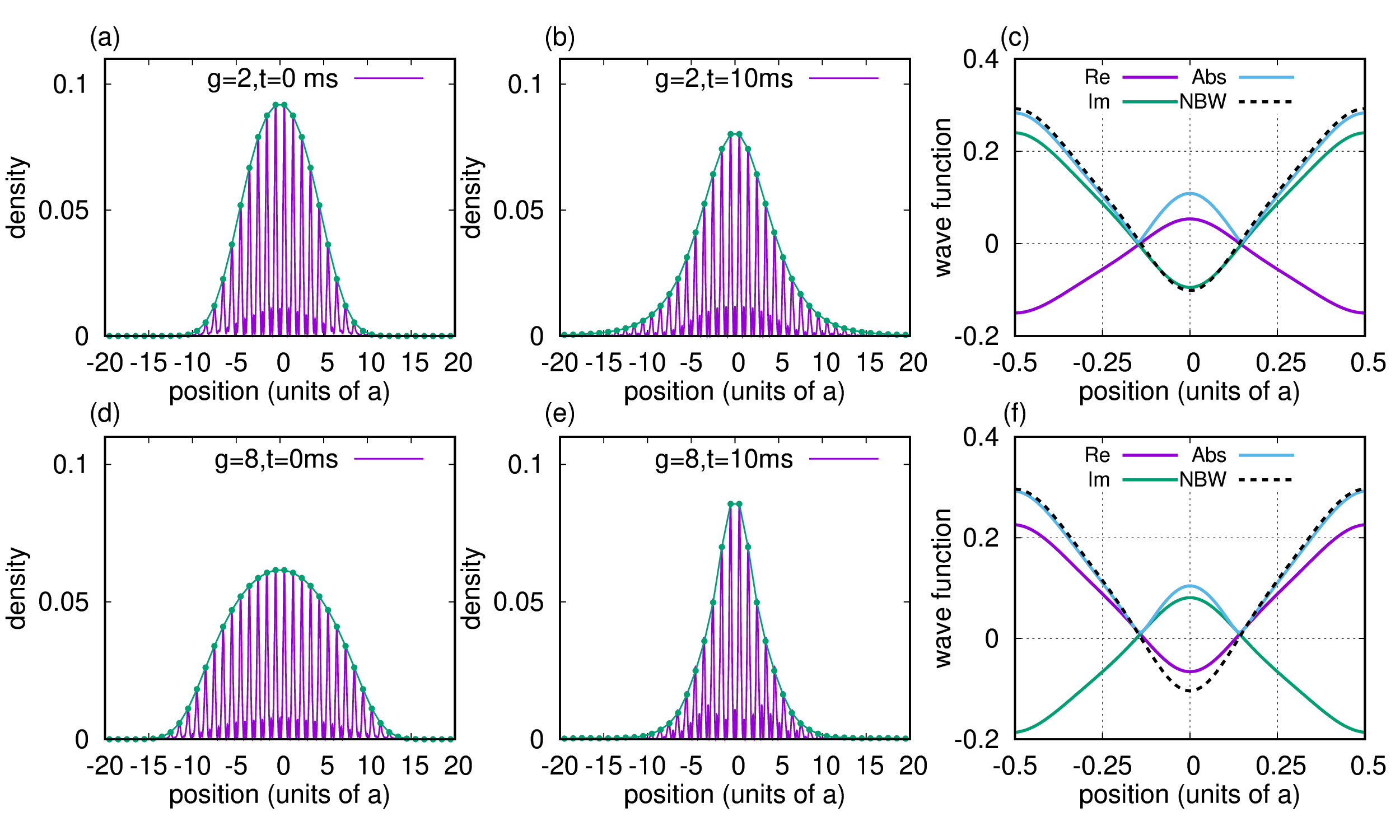}
 \end{center}
 \caption{
The density profile immdeiately after the loading process for (a)$g=2$ and (c)$g=8$. 
Purple and green line corresponding to the density profile and envelope function in position space, respectively.
The envelopes are well fitted by the Gaussian function, reflecting the initial state.
(b) and (e) corresponding to the density profile after 10ms holding with $g=2$ and $g=8$, respectively.
The non-soliton component goes away from the origin; thus, only the soliton solution remains around $x=0$.
The corresponding envelopes are hyperbolic secant in both cases.
(c) and (f) are enlarged views of the central site $-0.5 \le x \le 0.5$ of (b) and (e), respectively.
Purple, green, and light blue correspond to the real part, imaginary part, and absolute value of the wave function.
The black dashed lines indicate the NBW $\phi_{NB}(1,0,x)$ of the 1st excited band at $q=0$ with (c)$g_{lat}=0.06$ and (f)0.16.
 }
 \label{fig:env}
\end{figure}

In order to show the solitons more clearly, we plotted the density profile after the loading process in Fig.\ref{fig:env}.
Fig.\ref{fig:env} (a) and (c) shows density profiles immediately after the loading for $g=2$ and $8$.
A Gaussian shape approximately describes the profile since the loading process is not long enough to modify the envelope.
After the 10ms holding time, the non-soliton component goes far enough away from the origin of the initial state, and the immobile component remains at $x=0$.
Then the envelope of the wave packet becomes hyperbolic secant((b) and (d)), which is one of the characteristic behavior of solitons.
In addition to that, in a smaller structure, the wave functions have two nodes per site((c) and (f)).
This structure suggests that the soliton solutions are attributed to the NBWs in the 1st excited band.
We estimated the averaged chemical potential as $\braket{\mu(t)}=\int_{x_{l}}^{x_{u}} \psi(x,t)^\ast \mu \psi(x,t) dx/ \int_{x_{l}}^{x_{u}} \psi(x,t)^\ast \psi(x,t) dx$.
We chose $x_{l}=-20\pi$ and $x_{u}=20\pi$ for numerical convergence.
The chemical potential oscillates as a function of time since the wave packet is in breathing mode.
Therefore, we showed the minimum value of the averaged chemical potential between 8-12 ms $\braket{\mu}_{min}$ in Table.\ref{tab:CPP}.
The averaged chemical potential slightly lower than the chemical potential, which is estimated by the mean density immediately after the loading process.
However, the chemical potential lies between the 1st and 2nd excited linear bandgap(4.58$\le \mu \le$5.19), although it is difficult to identify the soliton solution exactly.
We also showed $P_{sol}(t)=\int_{x_{l}}^{x_{u}} | \psi(x,t) |^2 dx$ to estimate the population of the soliton at t=12ms.
As the interaction becomes stronger, the population of the soliton decreases.
The details of the soliton solution in the bichromatic OL system, such as spectrum, fundamental gap soliton, are interesting, but the discussion is beyond the scope of this paper.
In passing, we note that it is difficult to generate specific solitons or to specify soliton modes at this stage since the final state is sensitive to the initial state, time-sequence, and so on.

\begin{table}[h]
\caption{
Relationship between the interaction strength $g$ and the chemical potential of the solitons $\braket{\mu}_{min}$.
The chemical potential $\mu(q=0)$ of the 1st excited band at $q=0$ is obtained by Fig.\ref{fig:LBS}(b) and (c). 
The wavefunction of the central site immediately after the loading has a high density because of the extra components.
$\mu(q=0)$ is always higher than $\braket{\mu}_{min}$.
We numerically estimated the soliton population $P_{sol}$ at t=12ms for future experiments.
See text for more detail.
}
\begin{tabular}{c|c|c|c}
\hline\hline
  $g$  &  $\mu(q=0)$ & $\braket{\mu}_{min}$  & $P_{sol}$      \\
\hline
 2     &    4.71     &    4.61               &  0.902       \\
 4     &    4.78     &    4.68               &  0.878       \\
 6     &    4.85     &    4.74               &  0.819       \\
 8     &    4.90     &    4.79               &  0.705       \\
\hline\hline
\end{tabular}
\label{tab:CPP}
\end{table}

\begin{figure}[htbp]
 \begin{center}
 \includegraphics[width=8cm]{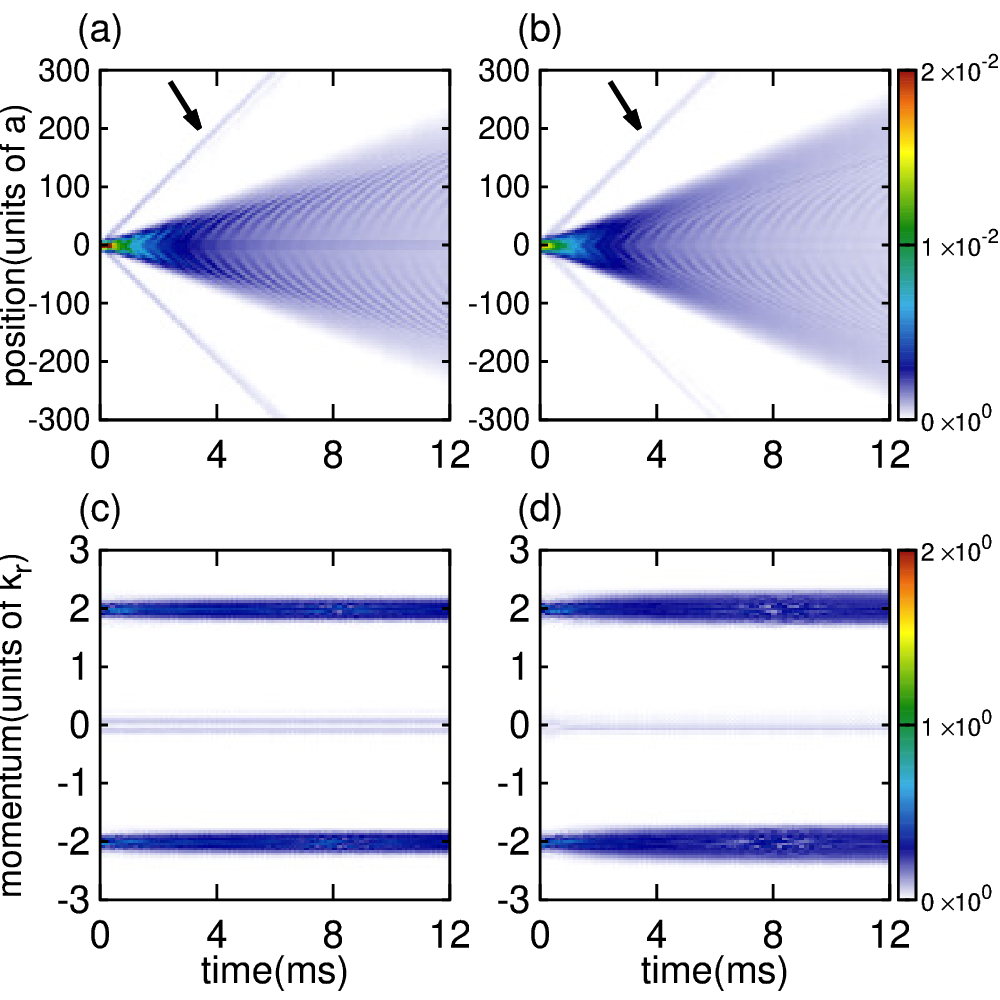}
 \end{center}
 \caption{
 Time evolution of the wave packet in position space((a) and (b)) and momentum space((c) and (d)) with the protocol for loading atoms onto anti-symmetric solutions.
Here the target state is the bottom of the 2nd excited band.
(a) and (c) corresponding to weakly interacting case with $g=1$.
(b) and (d) corresponding to strongly interacting case with $g=8$.
The interaction strength does not drastically alter the trends.
In both cases, fast-moving wave packets appear in position space(black arrow).
These wave packets are consist of 4th excited band components, unlike the symmetric loading case(Fig.\ref{fig:S8-DENSITY}).
 }
 \label{fig:S8-ANTISYM}
\end{figure}

We also examined the loading process to the bottom of the 2nd excited band(even to odd) with $s_1=10,s_2=8$.
Fig.\ref{fig:S8-ANTISYM} shows the time evolution of the wave packet with (a)$g=1$ and (b)$8$.
In position space, the loaded atoms spread out over time, same as the Fig.\ref{fig:S8-DENSITY}(a).
As seen in Fig.\ref{fig:LBS}(c), the 2nd excited bands are not strongly deformed by the interaction, and the gap appears around $|q|=1$.
Therefore, in contrast to the previous case, the post-loading dynamics behave as in the linear case (or weakly interacting case) even when the interaction is strong $g=8$.
The momentum distribution shows that most of the atoms appear around $|p|=2$, which means the protocol successfully loaded atoms onto the bottom of the 2nd excited band.
We note that, as discussed in Ref.\cite{Beijing-5}, the OL pulses with non-zero relative phases $\phi_{11}=\phi_{21}=\phi_{12}=\phi_{22}\ne0$ reduces the loading efficiency, small portions are loaded onto the ground and the 4th excited band.
Table.\ref{tab:POP-AS} shows the loading rate $R_{m}$.
In both cases, 15\% of atoms are loaded onto the 4th band-edge.
In order to achieve higher loading efficiency, an optimization based on TD-GPE would be necessary.

\begin{table}[h]
\caption{
The loading rate $R_{m}$ with $g=$1 and 8.
Same as Table.\ref{tab:POP}, $R_{1}$, $R_{3}$ are always smaller than $10^{-3}$.
}
\begin{tabular}{c|c|c}
\hline\hline
           &    $g=1$    &    $g=8$           \\
\hline
 $R_0$     &    0.050    &    0.050           \\
 $R_2$     &    0.714    &    0.748           \\
 $R_4$     &    0.179    &    0.152           \\
\hline\hline
\end{tabular}
\label{tab:POP-AS}
\end{table}

In passing, we note that we have examined the loading process with a standard band of $s_1=10,s_2=5$.
In this case, the atoms are loaded onto the soliton solution between the 1st and 2nd excited bands with the anti-symmetric(even to odd) loading protocol.
Appendix.\ref{sect:normal-band} discusses more detail of the results with the normal band.
We also note that our procedure is valid in the limit of monochromatic lattice, e.g., $s_1=10,s_2=0$(standard band).
Although anti-symmetric loading is required for soliton generation, the experimental implementation may be simpler than the bichromatic system.
The loading process with the Dirac-like point is of sufficient interest that it will be studied separately elsewhere.

\subsection{Case 2: Post-excitation dynamics with harmonic OL}
\label{sect:harmonic}

\begin{figure}[htbp]
 \begin{center}
 \includegraphics[width=5cm]{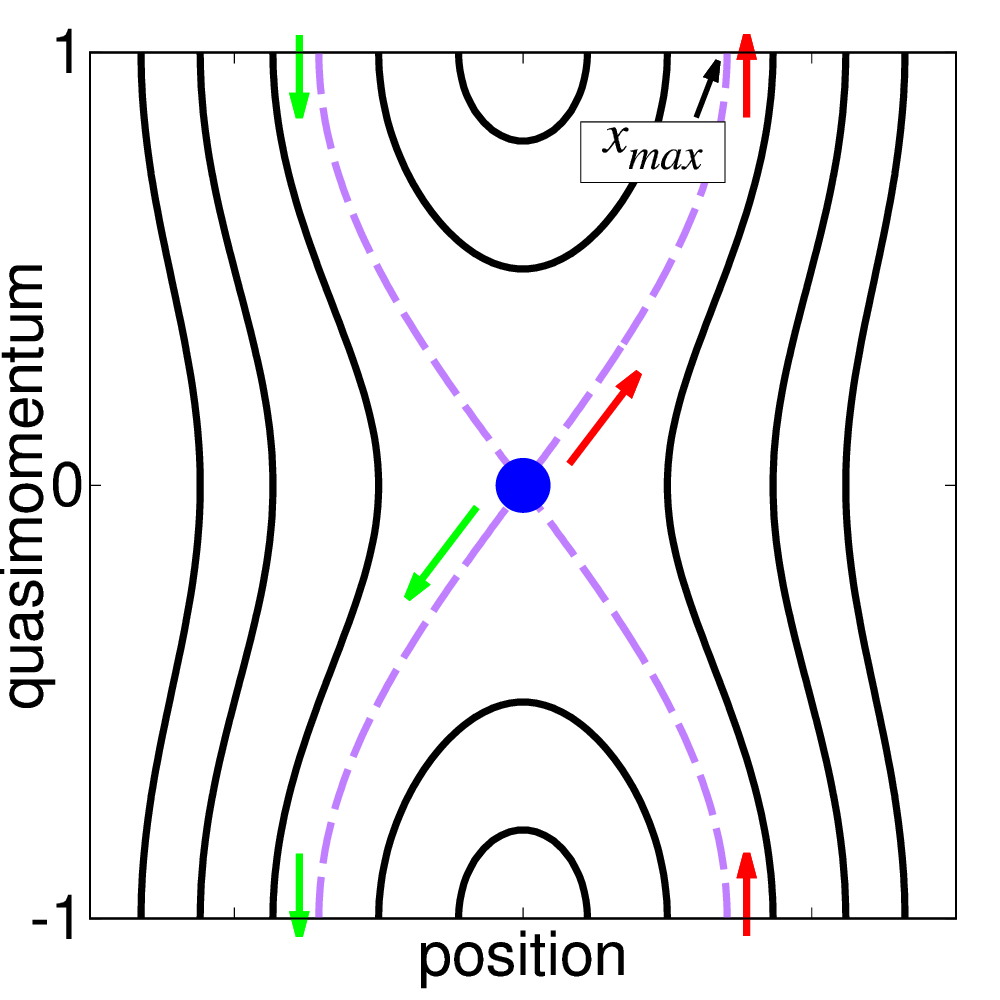}
 \end{center}
 \caption{
 Typical isoenergetic surface of the 1st excited band with the reduced zone representation.
 The surface is given by the 1-dimensional classical pendulum $H_{cl}$.
 The purple dashed line and the blue circle corresponding to a separatrix and a rough coordinate of the atoms immediately after the loading.
 In our case, the atoms are loaded onto the origin of the surface, then half of the wave packet moves up along the separatrix lines (red arrow), and the rest moves down(green arrow).
 Due to the Bragg reflection, the quasimomentum of the wave packet is reversed when they reach the band edge.
 }
 \label{fig:ECS}
\end{figure}

In the ultracold atomic system, time modulation of the optical lattice and time variation of the external potential can be experimentally realized\cite{eckardt}.
Band spectroscopy of the atoms in OL is one of the essential applications.
By combining linear potentials to OL, a St\"uckelberg-type interferometer can be constructed and measured by observing momentum distributions\cite{stuckel}.
Dispersions can also be measured by modulating the height of the optical lattice in time to inducing interband transitions\cite{spectro}.
Aarhus group experimentally measured the spectrum formed by the combined potential of the OL and the harmonic potential by using amplitude modulation of the OL with the acceleration of atoms due to the harmonic potential\cite{Aarhus2}.
Thus, the observation of the dynamics of atoms in the combined potential of the OL and the external potential is important for further applications.
Here we discuss the dynamics of loaded atoms with strong nonlinearity under the influence of the harmonic trap to identify global band dispersion and stability.
From the viewpoint of the classical interpretation under the single-band approximation, the dynamics is simply given by $H_{cl}(q,x)=\mu(q)+\nu x^2$\cite{Hamburg}. 
Fig.\ref{fig:ECS} shows a typical isoenergetic surface of the 1st excited band in the linear limit($g=0$) with a reduced zone representation.
The dispersion can be treated as a cosine function under the 1st-order tight-binding approximation in the linear limit.
Therefore, a trajectory of the loaded atoms traces a classical pendulum\cite{my-1,my-2,my-4}.
If the atoms are loaded on the origin of the surface, the atoms follow a line of the separatrix.
According to $H_{cl}$, we can easily estimate the maximum position of the atoms $x_{max}$.
In the case of $s_1=10,s_2=8$ and $\nu=1.2\times10^{-4}$, $x_{max}\simeq$ 33 sites(see Ref's.\cite{my-1,my-4}for the detail.)
Since the small nonlinear interaction does not alter the band dispersion drastically, the classical interpretation may work even with the nonlinear term $g\leq 8$.
Thus, measuring post-loading dynamics allows us to compute a general form of the band where the atoms are loaded.

\begin{figure}[htbp]
 \begin{center}
 \includegraphics[width=8cm]{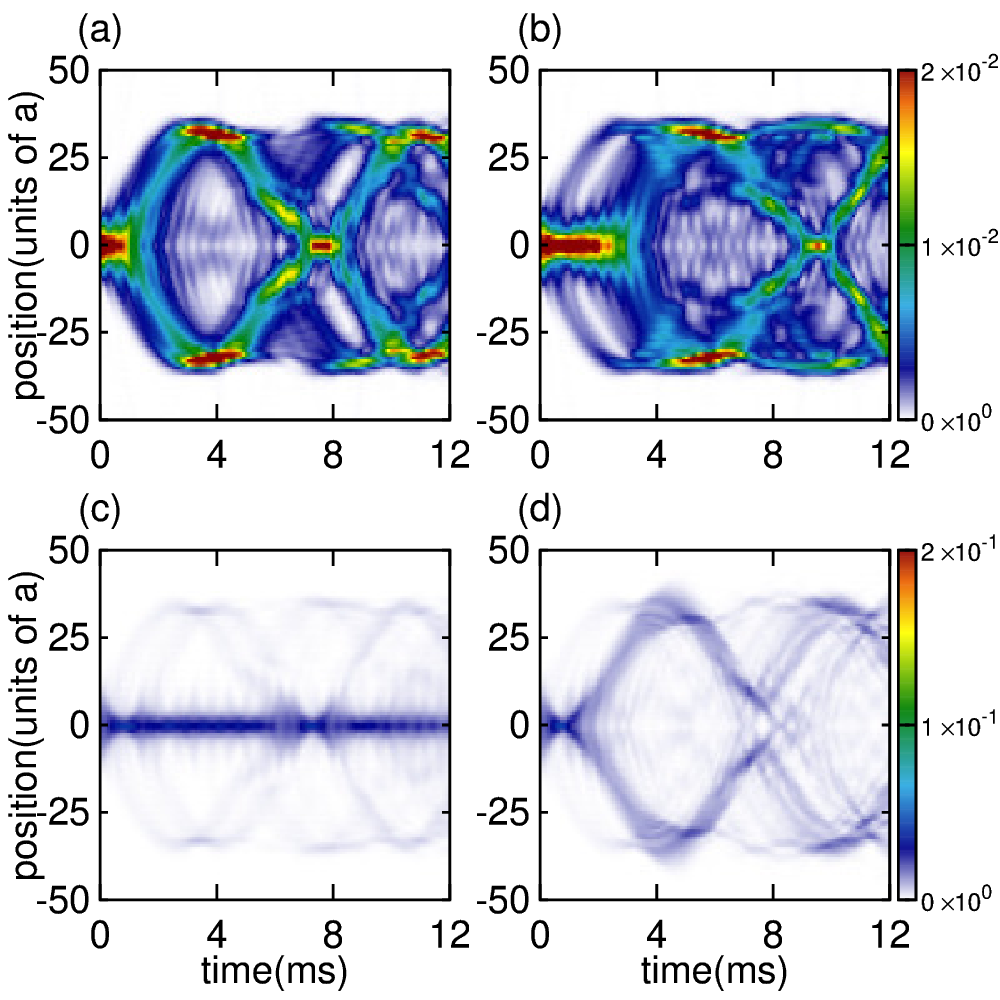}
 \end{center}
 \caption{Time evolution of the wave packet in position space as a function of holding time with $s_2=8$ for (a)$g=2$, (b)$g=4$, (c)$g=6$ and (d)$g=8$.
 Atoms are initially loaded to soliton solutions and the soliton decays to 1st excited band after a certain holding time.
 The atoms follow the isoenergetic surface of the 1-dimensional classical pendulum given by $H_{cl}$(see Fig.\ref{fig:ECS}).
 }
 \label{fig:S8-HT-D}
\end{figure}

\begin{figure}[htbp]
 \begin{center}
 \includegraphics[width=8cm]{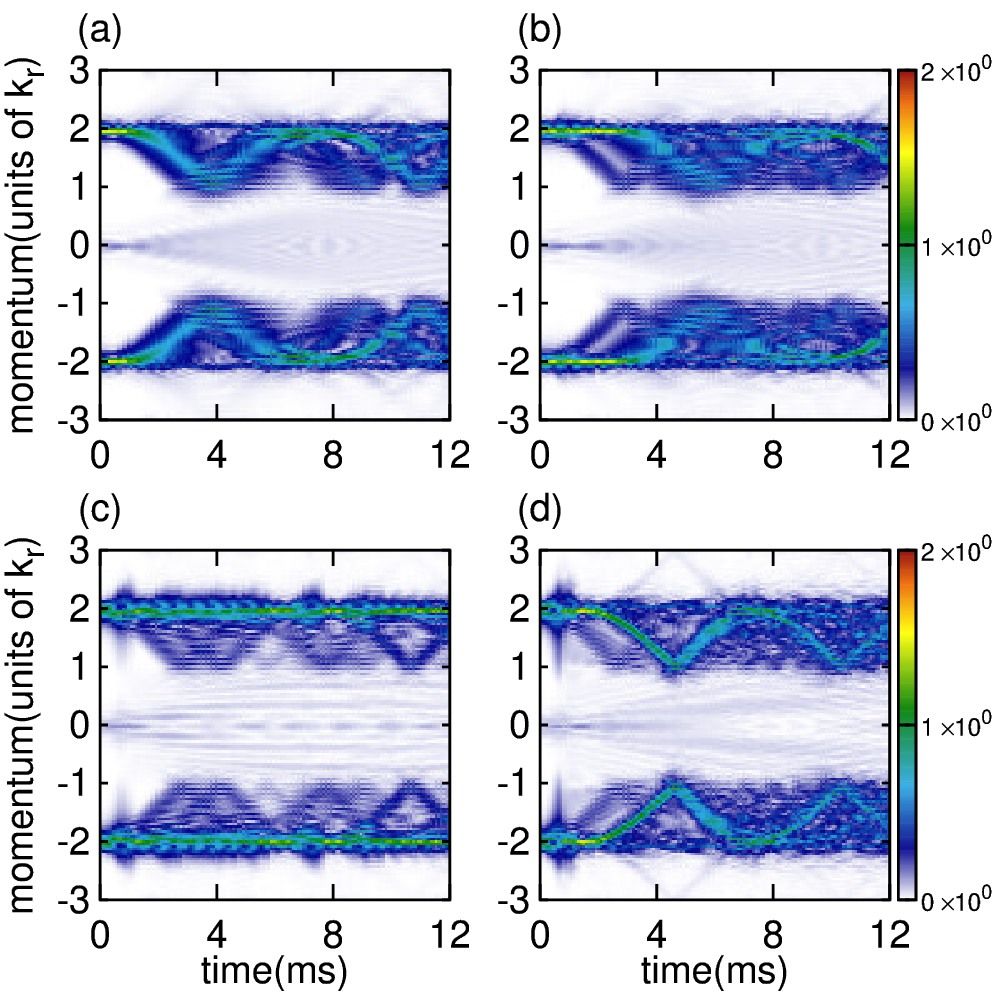}
 \end{center}
 \caption{Time evolution of the wave packet in momentum space after the band mapping process as a function of holding time with $s_2=8$ for (a)$g=2$, (b)$g=4$, (c)$g=6$ and (d)$g=8$ corresponding to Fig.\ref{fig:S8-HT-D}.
 }
 \label{fig:S8-HT-M}
\end{figure}

\begin{figure}[htbp]
 \begin{center}
 \includegraphics[width=8cm]{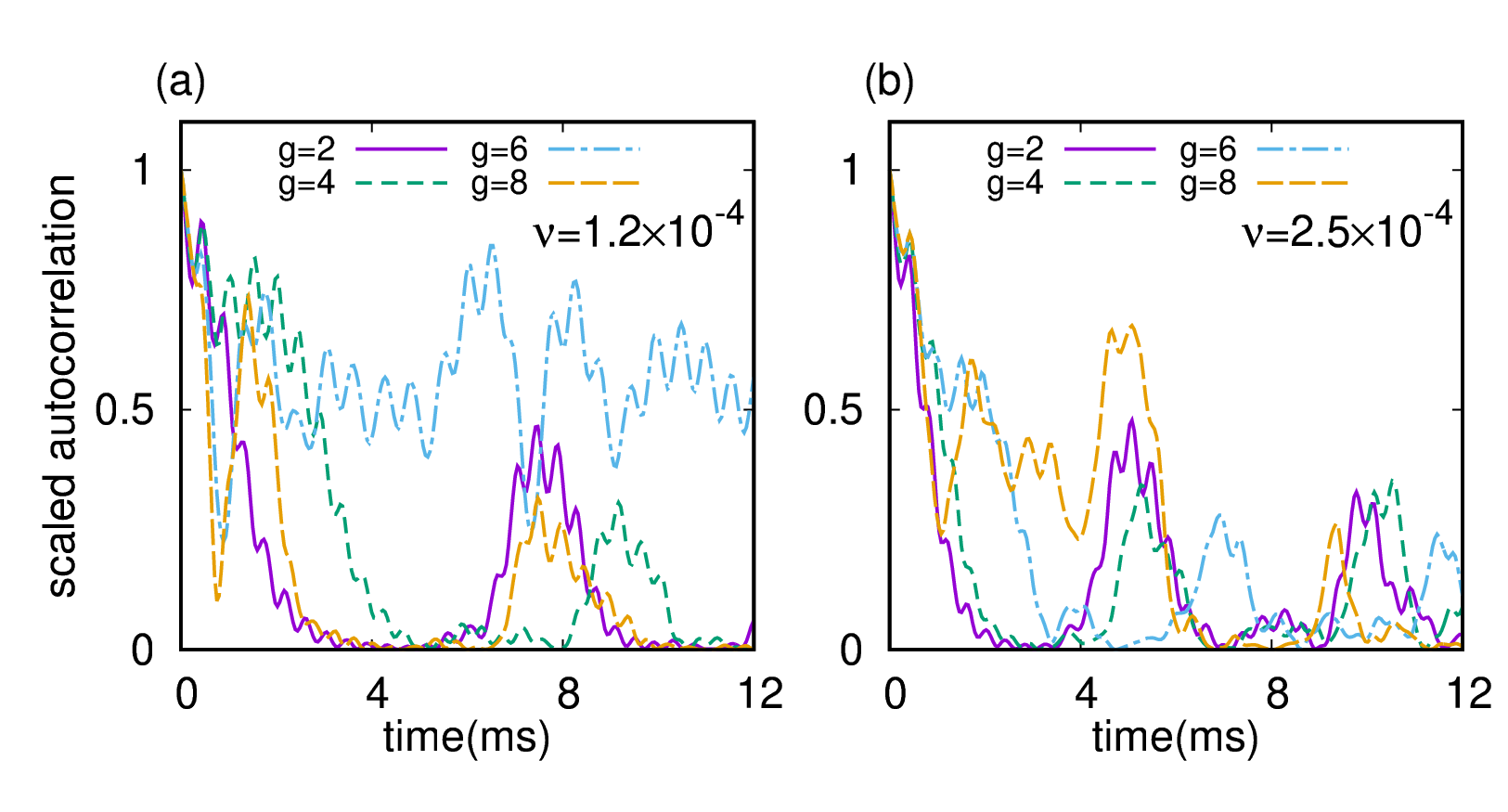}
 \end{center}
 \caption{
 Autocorrelation function $A(t)$ as a function of holding time with (a)$\nu=1.2\times10^{-4}$ and (b)$\nu=2.5\times10^{-4}$.
 Purple solid, green dashed, light blue chain and orange long-dashed corresponding to $g=2$, 4, 6, and 8, respectively.
 (a)In the case of $g=6$, the autocorrelation shows a stable feature.
 In other cases, the soliton decays within 4ms; subsequently, it shows revivals due to the closed classical trajectories given by $H_{cl}$.
 (b)In the case of (b)$\nu=2.5\times10^{-4}$, $g=8$ is most stable.
 See text for more detail.
 }
 \label{fig:AUTO-HT}
\end{figure}

Fig.\ref{fig:S8-HT-D} shows the post-loading dynamics of the symmetric process with harmonic trap for $s_1=10,s_2=8$ with $\nu=1.2\times10^{-4}$.
Fig.\ref{fig:S8-HT-M} shows corresponding momentum distributions after the band mapping process.
As shown in Fig.\ref{fig:S8-DENSITY}, the process loads atoms to soliton solution in $g=2-8$.
The acceleration force due to the harmonic potential lowers the density of the soliton so that the loaded atoms decay to the 1st excited band from the soliton solution after a certain holding time.
In terms of the tight-binding approach, the harmonic potential modifies the spatially uniform tunneling constant to nonuniform.
Therefore, in the presence of the harmonic trap, the soliton solution of the uniform lattice is not stable anymore.
As increasing the interaction strength, the soliton gets more stable up to $g=6$ which corresponds to $g_{lat}=0.13$(See Fig.\ref{fig:LBS} (b)).
When the interaction strength reaches $g=8$, the soliton decays immediately after the loading.
To confirm this effect more clearly, we also plot a scaled autocorrelation function
$$
A(t)=|\braket{\Psi(t=\tau_{total})|\Psi(t)}|^2/|\braket{\Psi(t=\tau_{total})|\Psi(t=\tau_{total})}|^2
$$
as a function of the holding time in Fig.\ref{fig:AUTO-HT} with (a)$\nu=1.2\times10^{-4}$ and (b)$\nu=2.5\times10^{-4}$.
$\Psi(t=\tau_{total})$ is a wavefunction immedeately after the loading process.
In (a), the autocorrelation function is most stable at g=6.
In the other cases, the autocorrelation functions quickly become unstable, then show revival due to the classical orbit given by $H_{cl}$. 
We note that this phenomenon depends on the balance between the nonlinear term and the strength of the harmonic trap.
In the case with (b)$\nu=2.5\times10^{-4}$, the autocorrelation function shows higher stability for $g=8$ than for $g=6$.
The effects due to the external potentials are interesting because they are related to manipulating solitons in the coordinate space, but we leave it out in this paper.

\subsection{Case 3: Effect of Instability}
\label{sect:instability}
One of the unique features of the lattice BEC system is dynamical instability.
This instability is considered a counterpart of an energetic instability and causes NBWs(solitons) to be unstable when additional perturbative modes exist.
To explore dynamical instability of NBWs, we follow the discussion in Refs.\cite{NLBloch1,NLBloch2}.
We assume that NBW experience a small perturbation with momentum $f$ as $\psi=\phi_{NB}(n,q,x)+\delta\psi$, where $\delta\psi=e^{iqx}\{ u_f (x) e^{ifx} + v_f^{\ast} (x) e^{-ifx} \}$.

The perturbative part $\delta\psi$ obeys the time-dependent equation
\begin{eqnarray}
\frac{d}{dt} \left( \begin{array}{cc} u_f \\ v_f \\ \end{array} \right) = \hat{\sigma_z} \hat{B} \left( \begin{array}{cc} u_f \\ v_f \\ \end{array} \right) = \Lambda_l(q) \left( \begin{array}{cc} u_f \\ v_f \\ \end{array} \right)
\label{eq:pertur}
\end{eqnarray}
where the operator
\begin{eqnarray} 
\hat{B} = \left(\begin{array}{cc} L_+ & g_{lat}\Phi^2(n,q) \\ g_{lat} \left\{\Phi^{\ast}(n,q)\right\}^2 & L_- \\ \end{array} \right)
\label{eq:operator}
\end{eqnarray}
and diagonal part
\begin{eqnarray}
L_{\pm} = - \left[ \frac{d}{dx} + i (\pm q+f)  \right]^2 + s_1 \sin^2(x) +s_2 \sin^2(2x) - \mu +2g_{lat} |\phi_{NB}(n,q,x)|^2 
\label{eq:diagonal}
\end{eqnarray}
with Pauli matrix $\hat{\sigma_z} = \left(\begin{array}{cc} 1 & 0 \\ 0 & -1 \\ \end{array} \right)$ and $\Phi(n,q)=e^{-iqx} \phi_{NB}(n,q,x)$.
Here the operator $\hat{\sigma_z} \hat{B}$ is non-Hermitian; thus, eigenvalues $\Lambda_l(q)$ can be complex.
Besides, complex eigenvalues always appear in pairs, one has a positive imaginary part, and the other has a negative.
Therefore, if the operator $\hat{\sigma_z} \hat{B}$ has complex eigenvalues, the state grows exponentially in time.

\begin{figure}[htbp]
 \begin{center}
 \includegraphics[width=8cm]{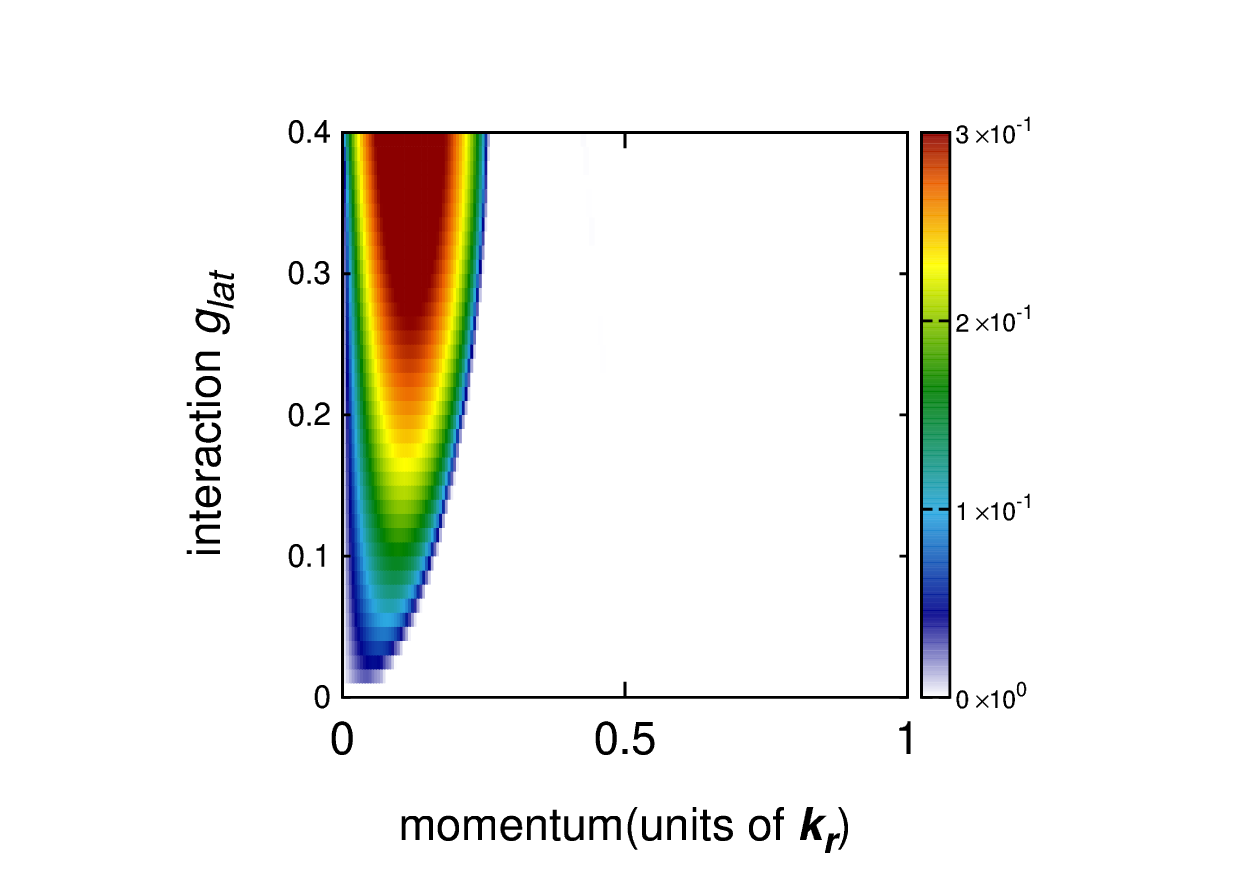}
 \end{center}
 \caption{
 The maximum absolute values of the imaginary part of the eigenvalues $\Im \Lambda_l(q)$ for the 1st excited band.
 The unit of the colorbar is $E_r$.
Vertical axis corresponds to interaction strength $g_{lat}$ and horixontal to perturbative momentum $f$.
 Value becomes higher toward red and lower toward white.
 White corresponds to a dynamically stable regime.
The unstable regime appears around $f=0$.
 }
 \label{fig:STABILITY}
\end{figure}

Fig.\ref{fig:STABILITY} shows the maximum absolute values of the imaginary part of the eigenvalues $\Im \Lambda_l(q)$ of the 1st excited band at $q=0$.
A strongly unstable regime appears around $q=0$.
The result implies that the soliton solutions containing NBW of the 1st excited band at q=0 are unstable, according to the composite relation\cite{NLB-Soliton1}.
While features of dynamical instability can not be clearly seen within 12ms post-loading dynamics in the previous section, however, dynamical instability may break the stable wave packet for long-term propagation.

\begin{figure}[htbp]
 \begin{center}
 \includegraphics[width=8cm]{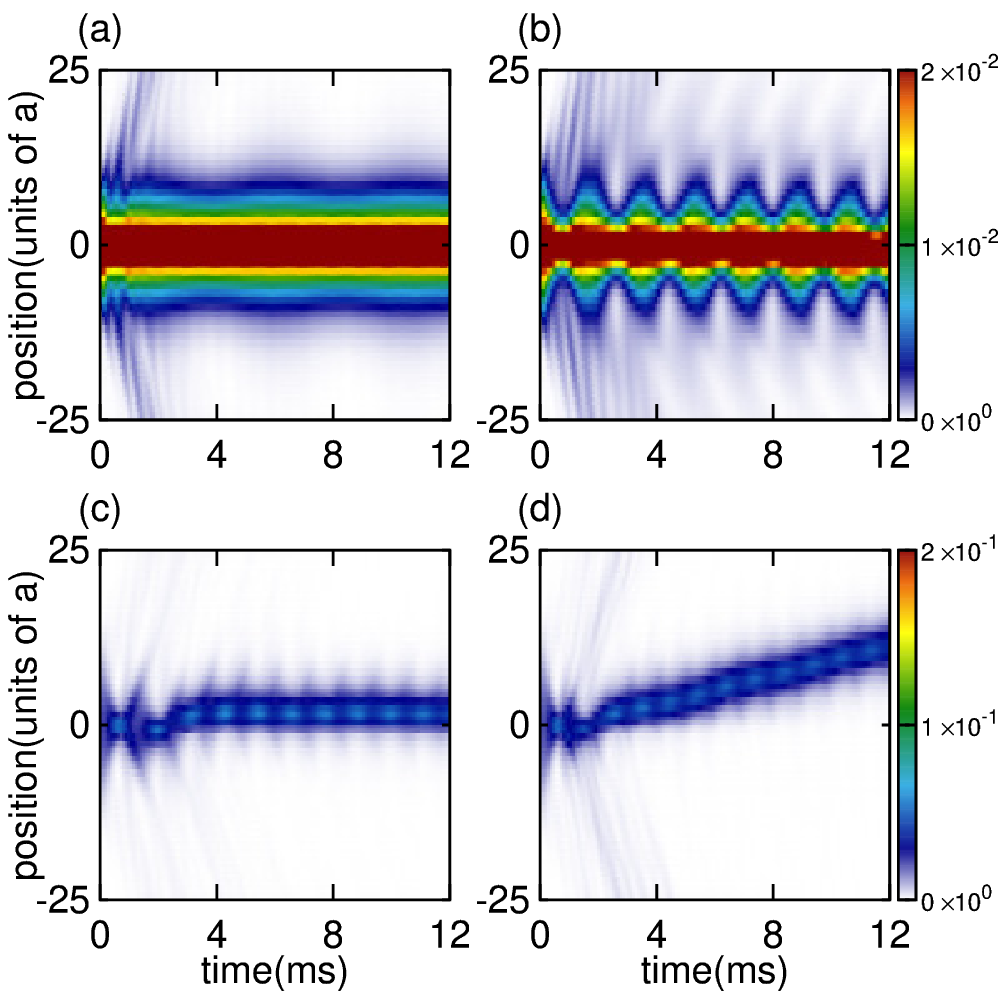}
 \end{center}
 \caption{Post-loading dynamics with a gaussian noise for $s_1=10,s_2=8$.
 In (a)$g=2$ and (b)$g=4$, the dynamics is stable despite the noise is added.
 On the contrary, in (c)$g=6$ and (d)$g=8$, the symmetry of the wave packet is broken due to the noise.
 }
 \label{fig:RND}
\end{figure}

\begin{figure}[htbp]
 \begin{center}
 \includegraphics[width=7cm]{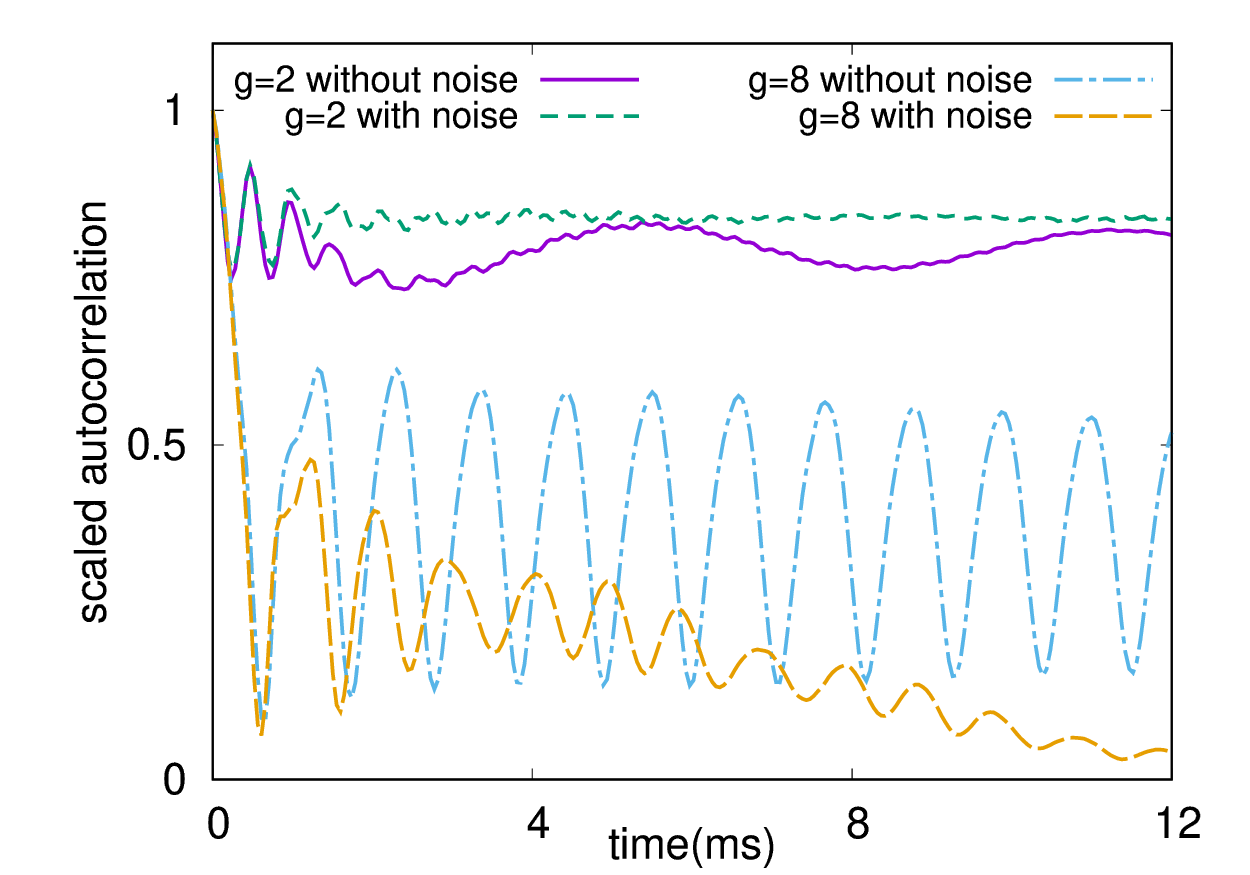}
 \end{center}
 \caption{ Autocorrelation function $A(t)$ as a function of holding time. 
 Green dashed (orange long-dashed) and purple solid (light blue chain) correspond to $g=2$($g=8$) with and without noise.
 }
 \label{fig:AUTO}
\end{figure}

In order to investigate dynamical instability more deeply, we numerically solved TD-GPE with an artificial gaussian type noise in addition to the condensate initial function $\Psi(t=0,x)=\psi(x)+\Delta(x)$(see Ref.\cite{noise}), namely,
\begin{eqnarray}
\Delta(x)=\epsilon \psi(x=0) r(x) e^{-x^2/\sigma^2}
\label{eq:Delta}
\end{eqnarray}
where $\psi(x=0)$ is the BEC in the harmonic potential and $r(x)$ is a random number uniformly distributed in $[0,1]$.
We chose $\epsilon=0.1$ and $\sigma=20\pi$.
Fig.\ref{fig:RND} shows the density profile after the loading process as a function of holding time in position space with $s_2=8$.
Only the OL is turned on while the holding process. 
In (a) and (b), dynamical instability does not strongly break the symmetry within 12 ms, although the imaginary part appears around $f=0$.
The absolute value of the imaginary part increases as it increases the interaction strength to $g=6$, the effect of dynamical instability can be seen clearly in the density profile.
We also plot the scaled autocorrelation $A(t)$ as a function of holding time in Fig.\ref{fig:AUTO}.
In the case of $g=2$, the noise makes the autocorrelation stable since it suppresses a coherence of the breathing mode.
We note that the random noise might correspond to experimental imperfections, e.g., unspecifiable excited state; however, this analysis is more of a numerical check.
The shape of the noise function may be spatially correlated, and the random noise may be totally different from the real experimental environment.
It is required to know the sources of the noises more precisely to investigate the environmental effect.

\section{Conclusions}
\label{sect:conclusions}

We discussed the standing-wave pulse sequence\cite{Beijing-1,Beijing-5} in the 1-dimensional bichromatic system with atom-atom interaction in terms of mean-field theory.
In the previous paper\cite{my-4}, we showed that the pulse sequence is essentially multi-path interference in the momentum domain.
And the parity selection rule of the pulse sequence is specified by the orthonormal Bloch solutions in the linear case.
We also verified that the weak nonlinearity $g<1$ does not alter the results dramatically.
In this paper, we extended our theoretical work\cite{my-4} to the strongly nonlinear regime.
Contrary to the linear case, the NBWs are non-orthonormal, and additional loop structures appear at the band edges in the time-independent spectrum.
In addition to that, the spatially localized soliton solutions appear, which are connected to NBWs through a composite relation\cite{NLB-Soliton1}.
Therefore, the loading process to NBWs is a nontrivial issue.

In order to examine the loading process to NBWs, we numerically simulated the pulse sequence with the inverted band.
We found that the sequence can load atoms to the soliton solution in the excited band with an appropriate time interval of the pulses when the interaction strength is high enough.
And we numerically verified that the sequence is valid up to $g=8$.
The post-loading dynamics with only OL show that the non-soliton components quickly move away, leaving the soliton solution at the initial position of the wave packet.
In the periodic potential, the soliton component stays in its initial position and oscillates in the breathing mode according to the interaction strength.
Subsequently, we showed the post-loading dynamics with the addition of an external harmonic potential to check the solitons' stability.
The harmonic potential modulates the tunneling constant; thus, the soliton decays after a certain holding time.
Finally, we discussed the effect of the dynamical instability in terms of the linear stability analysis\cite{NLBloch1,NLBloch2}.
The spectrum of the time propagation operator shows that the nonlinear Bloch solutions are unstable at $q=0$.
In order to verify the dynamical instability, we numerically simulated noise by solving the TD-GPE using an artificial noise function mixed into the BEC's initial state and found that the process nevertheless succeeded in loading atoms to the soliton solution.
However, the noise suppresses coherence of the breathing oscillation and leads to instability while holding.
Our theoretical analysis recommends the range of interaction strength $g=2-6$ for future experiments.
This paper showed how to load atoms onto soliton solutions with the periodic potential.
The obtained soliton solutions are in the breathing mode with a hyperbolic-secant-type envelope.
The soliton solution has a complex spectrum that depends on many parameters\cite{NLB-Soliton1,soliton-the2}.
Therefore, the manipulation of the modes and the envelopes of solitons are expected to be difficult in the method of this paper.
Reinforcement learning using the nonlinear GP equation will be more appropriate than the brute force method with the linear Bloch equation.

Compared to the adiabatic process, the standing-wave pulse method can load atoms in a shorter period of time\cite{Beijing-1,Beijing-5}.
Moreover, compared to the other methods\cite{rapidload}, it only requires controlling time-intervals of lattice pulse with a precision of $0.1\mu$s.
However, there is still much to be explored from the theoretical point of view in the field of ultracold atomic physics.
For example, Ref.\cite{loop-exp} reports that there may be additional instability due to the beyond mean-field correlation.
The theoretical paper Ref.\cite{4th} considers 4th-order interaction proportional to $|\Psi|^4$ in the case of a strongly interacting limit.
A recent paper \cite{MODEM} took into account additional terms corresponding to 3rd-order correction ($\propto g_3$) and the three-body loss rate ($\propto K_3$) given in the form of $g_3 |\Psi|^3 - i K_3 |\Psi|^4$ which is called the Lee-Huang-Yang term.

As for future applications in the ultracold atomic physics, combinations of standing-wave pulse sequence and the other experimental techniques would lead to a powerful strategy for coherent manipulation of the nonlinear wave packets.
The amplitude modulation leads to coherent band couplings, which allow us to produce modified band structures with vast Hilbert space\cite{Holthaus,BF-trans}. 
In this paper, we applied the external harmonic potential for inducing intra-band transition as a very practical example.
Indeed, there are many other methods to induce intra-band transition.
As an example, a phase modulation technique such as the topological pumping process\cite{pump} would lead to unique coherent manipulations.
In addition to manipulations on the center-of-mass motion, manipulations on the internal degrees of freedom should be able to open up insights into designing new quantum devices.
Especially, the multi-component system would be a very nice platform.
The appropriate combinations of the coherent control may produce gap solitons with spin-orbit coupling\cite{Spin-orbit}, stabilized NBWs\cite{stab-fig}, and so forth.
Much remains to be studied.

\section*{Acknowledgments}
T.Y. acknowledges support from JSPS KAKENHI Grant Number JP20K15190.

\appendix

\section{Optimized time-interval for pulse-sequanece}
\label{sect:optimize}

The linear Bloch waves are given in the same formula as $\phi_B(n,q,x)=\sum_K C_B(n,q,K)e^{2iKx}$  which is the eigensolutions  of the time-independent Bloch Hamiltonian $H_B=-\frac{d^2}{dx^2}+s_1 \sin^2(x+\phi_1)+s_2 \sin^2(2\{x+\phi_2\})$.
Here $\phi_1$ and $\phi_2$ are controllable phases for the OL.
The Bloch coefficients $C_B(n,K,q)$ can be obtained by solving the recurrent formula
\begin{eqnarray}
(q+2K)^2 C_B (n,q,K) -s_2 e^{-4i\phi_2} C_B (n,q,K-2)/4 -s_1 e^{-2i\phi_1} C_B (n,q,K-1)/4 &&  \nonumber \\
 \nonumber
-s_1 e^{2i\phi_1} C_B (n,q,K+1)/4 -s_2 e^{4i\phi_2} C_B (n,q,K+2)/4 && \\
= (E_q^n-s_1/2-s_2/2) C_B (n,q,K), &&
\label{eq:rec}
\end{eqnarray}
namely, the central equation 
(see \cite{SSP} and our previous work \cite{my-4})
where $E_q^n$ represents the eigenenergy of the Bloch state.

In the linear regime, once the Bloch waves are obtained, the time-evolution of the two-step pulse sequence is given in the form of $\Psi(\tau_{total})=H_F(\tau'_2)H_B(\tau_2)H_F(\tau'_1)H_B(\tau_1)\Psi(0)$ where the Hamiltonian for free propagation $H_F=-\frac{d^2}{dx^2}$.
In order to optimize the pulse-sequence, we assume that the initial condition is the 0-momentum state, although the initial wave packet contains non-zero momentum states.
This approximation works well if the harmonic potential $\nu$ is loose and the interaction strength $g$ is high.
Since there is no accelerating force during the pulse sequence, we only consider the dynamics at $q=0$.
This assumption allows us to use a fidelity $F=|\braket{\phi_B(m,q=0)|\Psi(\tau_{total})}|^2$ as a informative index where $m$ is a target band index.
In this paper, our target states are the eigenstates of the Bloch Hamiltonian $H_B$ with the phases $\phi_1=\phi_2=0$.
During the on-duty cycle, we use the same lattice height $s_1,s_2$ for the target state; however, we vary $\phi_1, \phi_2$ for optimization.

As seen in Fig.\ref{fig:LBS}, the coefficients of the anti-symmetric solutions in $K$-space must have different signs for positive and negative indices $K$.
And the off-duty cycle with the free Hamiltonian $H_F$ gives a time-dependent phase change corresponding to the square of the momentum $e^{-4K^2 \tau'}$ for each $K$ momentum state.
Therefore, to achieve loading atoms onto anti-symmetric states, the on-duty cycle with $H_B$ needs to choose appropriate phases ${\phi_1,\phi_2}$.
For example, if we choose $\phi_1=\pi/4$ and $\phi_2=\pi/8$, the recurrent formula becomes
\begin{eqnarray}
(q+2K)^2 C_B (n,q,K) + i s_2 C_B (n,q,K-2)/4 + i s_1 C_B (n,q,K-1)/4 &&  \nonumber \\
 \nonumber
-i s_1 C_B (n,q,K+1)/4 -i s_2 C_B (n,q,K+2)/4 && \\
= (E_q^n-s_1/2-s_2/2) C_B (n,q,K). &&
\label{eq:rec-2}
\end{eqnarray} 
In this case, the on-duty cycle gives phases for momentum states with negative $K$, which is opposite to the phases for positive $K$ states.
Besides, we need to vanish the $K=0$ components using an interference effect since the $K=0$ components of the anti-symmetric state should be 0.
Thus, the effect of the OL phases on the optimization is nontrivial.
We numerically searched optimized time intervals by using a brute force method with a 0.1$\mu$s step size in each of the time duration $0 \le \tau_1, \tau'_1, \tau_2, \tau'_2 \le 40\mu s$.
We limit $\tau_{total} \le 100\mu s$ by following the previous studies\cite{Beijing-1,Beijing-2,Beijing-3,Beijing-5,my-4}.
To check the effect of the OL phases for loading atoms to the 2nd excited band(anti-symmetric solution), we chose 0, $\pi/4$ and $\pi/8$.
See Table.\ref{tab:phase-optimize-8} for the results.

The result shows that $\phi_{11}=\phi_{21}=\pi/4$ and $\phi_{12}=\phi_{22}=\pi/8$ is the best choice for loading atoms to the 2nd excited and the fidelity was found to be above 90\% for all combinations.
As for the experiments, it is better to choose the same phases $\phi_{11}=\phi_{12}=\phi_{21}=\phi_{22}$ to increase reproducibility.
So that, for simplicity, in this paper, we used $\phi_{11}=\phi_{12}=\phi_{21}=\phi_{22}=\pi/4$.
We note that we used 0-phase for the loading atoms to the 1st excited(symmetric solution) since the $\phi_{11}=\phi_{12}=\phi_{21}=\phi_{22}\ne0$ lowers the fidelity for loading.
As for the detailed discussions on the effects of the parabolic trap, see our previous paper\cite{my-4}.

\begin{table}[h]
\caption{
Optimized loading protocols to the first and the second excited bands with two step on- and off- procedure by using a brute force method with a period of 0.1$\mu$s grid\cite{my-4}.
The lattice height $s_1$ fixed to 10.
In this paper, we chose relative phase equals $\pi/4$ for loading atoms to the anti-symmetric solution.}
\begin{tabular}{cccc|cccc|c}
\hline\hline
  $\phi_{11}$  &  $\phi_{21}$  & $\phi_{12}$  & $\phi_{22}$  &  $\tau_1$  & $\tau'_1$  &  $\tau_2$    & $\tau'_2$  &  $F$        \\
\hline
 $\pi/4$       &    $\pi/4$    &   $\pi/4$    & $\pi/4$      &  33.8      &   34.7     &   14.8       &   4.5      &  0.939       \\
 $0$           &    $0$        &   $\pi/4$    & $\pi/4$      &  12.7      &   34.6     &   1.5        &   16.9     &  0.966       \\
 $\pi/4$       &    $\pi/8$    &   $\pi/4$    & $\pi/8$      &  32.8      &   39.9     &   39.9       &   6.5      &  0.967       \\
 $0$           &    $0$        &   $\pi/4$    & $\pi/8$      &  29.0      &   18.2     &   0.1        &   4.9      &  0.957       \\
\hline\hline
\end{tabular}
\label{tab:phase-optimize-8}
\end{table}

\section{Loading process with the normal band}
\label{sect:normal-band}

In this section, we briefly discuss a selection rule of the loading process with the normal band($s_1=10,s_2=5$).
Here, we numerically examine the loading process to the 1st excited(even to odd) and 2nd excited(even to even) band with $g=8$.

\begin{figure}[htbp]
 \begin{center}
 \includegraphics[width=8cm]{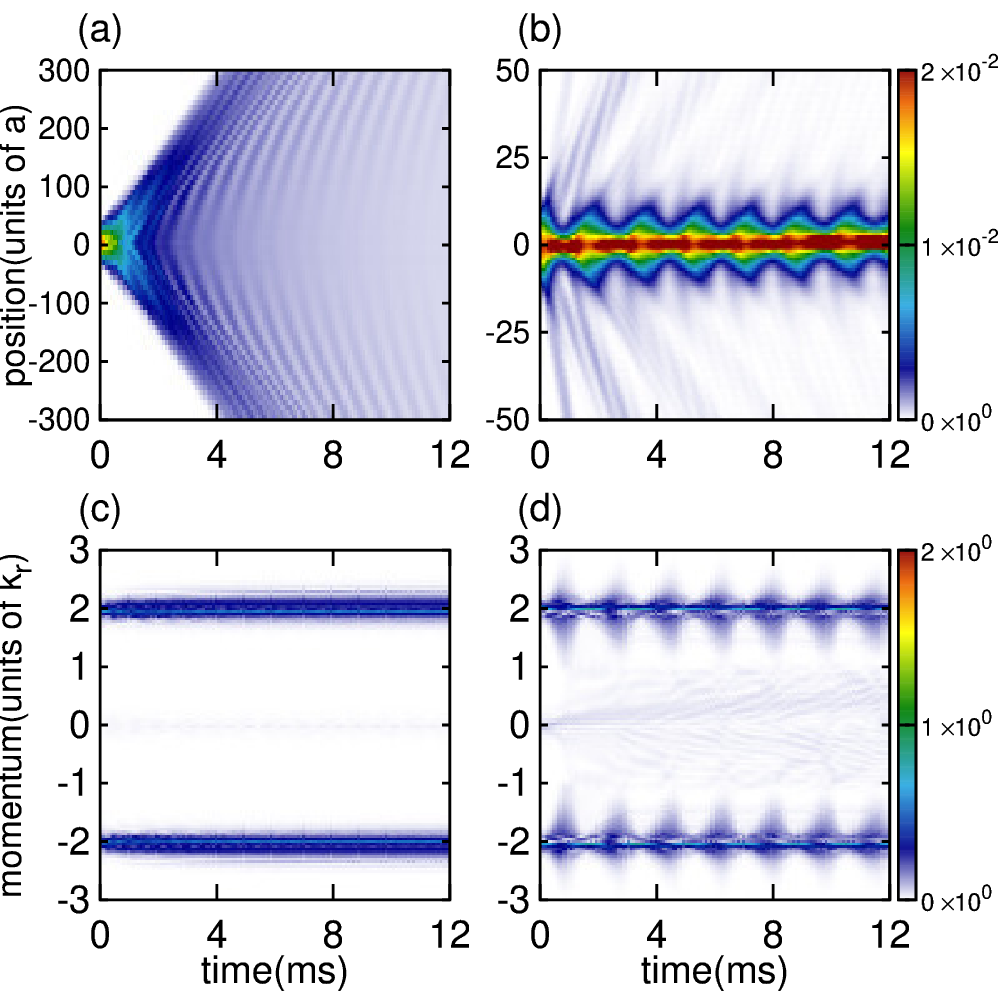}
 \end{center}
 \caption{ Time propagation of the wave packet as a function of holding time with the normal band $s_1=10,s_2=5$ for $g=8$.
 The optimized loading protocol is shown in Table.\ref{tab:s5}.
 (a) and (b) corresponding to the density distributions in position space with symmetric(even to even) and anti-symmetric(even to odd) loading process, respectively.
 (c) and (d) corresponding to the momentum distributions after the band mapping for (a) and (b).
 }
 \label{fig:s5}
\end{figure}

In the case of normal dispersion, at $q=0$, the 1st excited band solution is anti-symmetric, and the 2nd excited band is symmetric.
Therefore, in the linear limit, the loading process with the relative phases $\phi_{11}=\phi_{21}=\phi_{12}=\phi_{22}=0$ can load atoms to the 2nd excited band.
Fig.\ref{fig:s5} shows the time evolution immediately after the loading process.
Both of the process load atoms to around $|p|=2$ in momentum space((c) and (d)).
In position space, contrary to the case of the inverted band, the loading to the anti-symmetric solutions produces the solitonic state.
We also numerically confirmed that the trend is unchanged in the monochromatic lattice limit($s_2=0$, normal band).
The results for $s_1=10, s_2=0$ are not shown since it is similar to the case of $s_1=10,s_2=5$.
In this appendix, we just checked the validity of the pulse-sequence procedure.
Therefore, further research is needed on the properties of solitons in the normal band, such as dynamical instabilities.

\begin{table}[h]
\caption{
Optimized loading protocols with two step on- and off- procedure for $s_1=10,s_2=5$.
Same as Table.\ref{tab:two-col}, we chose relative phase equals to $\phi_{11}=\phi_{21}=\phi_{12}=\phi_{22}=\pi/4$ for anti-symmteric loading for on-duty cycles.
}
\begin{tabular}{c|cccc|c}
\hline\hline
  target   & $\tau_1$   &   $\tau'_1$       &  $\tau_2$    & $\tau'_2$  & F \\
\hline
2nd,even        &   3.8        &   4.9             &   23.1       &   28.9   &  0.999   \\
1st,odd         &   35.6       &   29.4            &   17.3       &   4.5    &  0.940   \\
\hline\hline
\end{tabular}
\label{tab:s5}
\end{table}

\section{Loop structure in bichromatic OL}
\label{sect:loop}
This appendix considers the loop formation in the band structure with the nonlinear bichromatic OL.
Here we use the lattice height $s_1=10,s_2=8$, same as the main discussion.

\begin{figure}[htbp]
 \begin{center}
 \includegraphics[width=8cm]{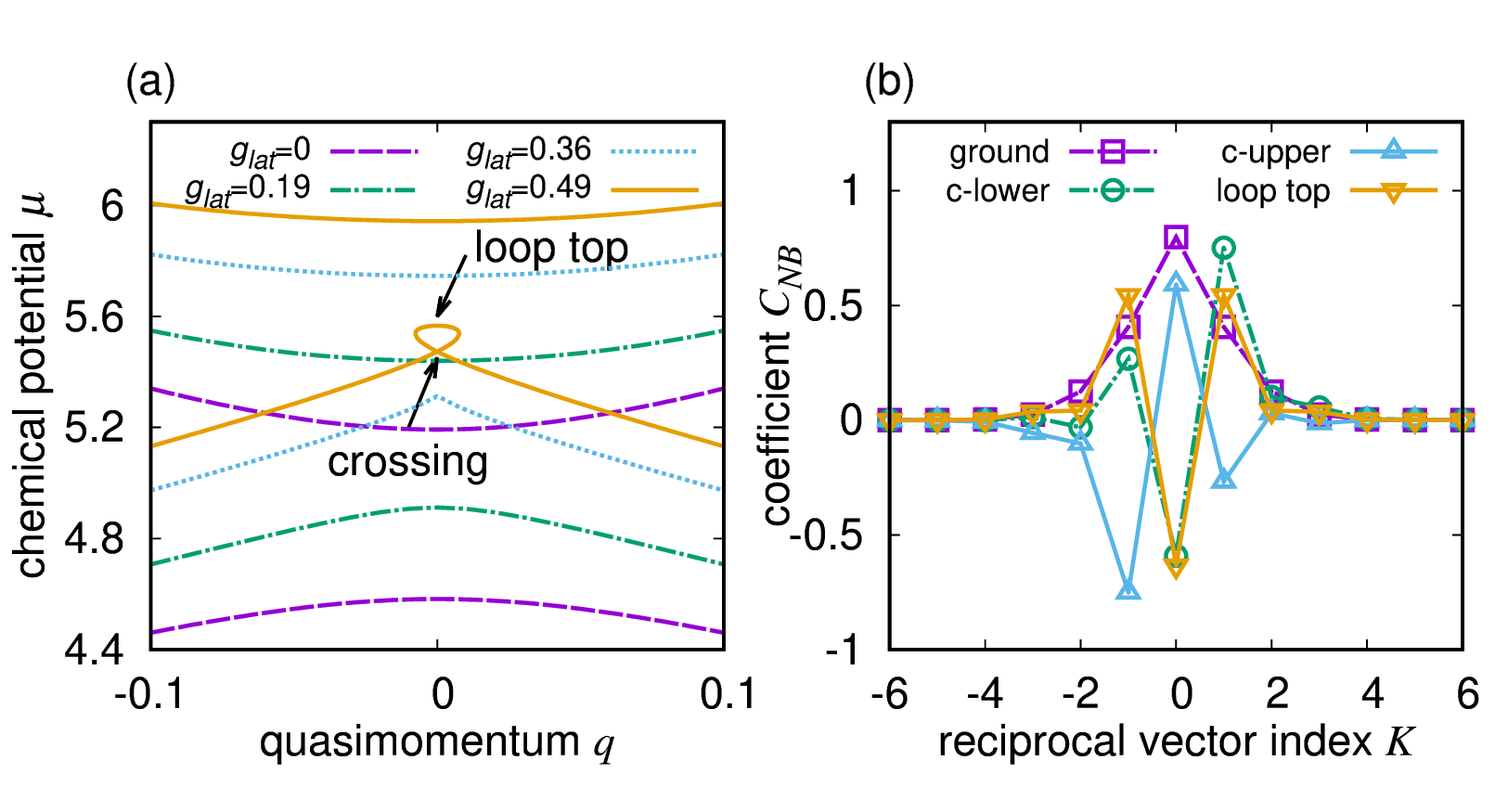}
 \end{center}
 \caption{
 (a) shows the band structure around 1st and 2nd excited bands.
Purple dashed, green chain, light blue short dashed and orange solid lines
correspond to $g_{lat}$ = 0, 0.19, 0.36 and 0.49. 
The interaction term gradually modifies the band structure, and the loop appears at a point a little short of $g_{lat}$ = 0.4.
(b) shows the first four Bloch coefficients for $g_{lat}$ = 0.49.
The coefficients at the crossing point(referred to as ``c-lower'' and ``c-upper'') are a superposition of an even and odd symmetry solution. 
The solution at the top of the loop has even symmetry.
 }
 \label{fig:loops}
\end{figure}

Fig.\ref{fig:loops} (a) shows the band structure around 1st and 2nd excited bands with $s_1 = 10, s_2 = 8$ for 4 values of $g_{lat}$.
As a precursor to the emergence of the loop, the derivative of the band dispersion $\frac{d\mu(q)}{dq} $ becomes discontinuous at $q=0$, if $g_{lat}=0.36$.
When $g_{lat}$ is greater than 0.4, the solution of the first excitation band at $q=0$ is split into three parts: two degenerate solutions and a loop top solution.
Then the loop appears at $q=0$ as seen in Fig.\ref{fig:LBS} (b).

Fig.\ref{fig:loops} (b) shows the first four NBW coefficients $C_{NB}(n,q,K)$ with $g_{lat}$=0.49.
The solution at the ground band has the same symmetry as in the
linear case(see Fig.\ref{fig:LBS}(d)). 
Two additional solutions at the crossing point consist of a superposition of odd and even symmetric solutions.
At the loop, the solution reflects the symmetry of the linear solution, which has a large overlap with the linear one.
We numerically checked that this trend remains unchanged up to $g_{lat}=1$(see Fig.\ref{fig:LBS}(a)).
This result suggests that it is possible to load atoms onto the loop top by the pulse-sequence method; however, it is necessary to discuss how the degenerate solutions affect the loading process.

\end{document}